\documentclass[iop]{emulateapj}
\usepackage{epsfig}
\usepackage{epstopdf}
\usepackage{natbib}
\usepackage{amssymb}
\usepackage{amsbsy}
\usepackage{natbib}
\usepackage{subfigure}
\usepackage[mathcal]{euscript}
\usepackage{float}
\usepackage{amsmath}
\usepackage{tabularx}
\usepackage{xspace}
\usepackage{enumitem}
\usepackage{color}

\usepackage{ulem,xspace}

\newcommand{\DPi}{$\Delta \Pi_1$\xspace}
\newcommand{\msun}{M$_\sun$}
\newcommand{\kms}{km~s$^{-1}$}
\newcommand{\vsini}{\ensuremath{v \sin{i}}}
\newcommand{\kepler}{\textit{Kepler}}

\bibpunct{(}{)}{;}{a}{}{,}

\begin{document}

\title{Core-Envelope Coupling in Intermediate-Mass Core-Helium Burning Stars}
\author{
Jamie Tayar\altaffilmark{1,2,3}, 
Paul G. Beck\altaffilmark{4,5,6},
Marc H.~Pinsonneault\altaffilmark{3}, 
Rafael A.~Garc{\'i}a\altaffilmark{7}, 
Savita Mathur\altaffilmark{5,6,8}
}
\altaffiltext{1}{Institute for Astronomy, University of Hawaii, 2680 Woodlawn Drive, Honolulu, Hawaii 96822, USA}
\altaffiltext{2}{Hubble Fellow}
\altaffiltext{3}{Department of Astronomy, Ohio State University, 140 W 18th Ave, OH 43210, USA}
\altaffiltext{4}{Institute of Physics, Karl-Franzens University of Graz, NAWI~Graz, Universit\"atsplatz 5/II, 8010 Graz, Austria.}
\altaffiltext{5}{Instituto de Astrof{\'i}sica de Canarias, E-38200 La Laguna, Tenerife, Spain}
\altaffiltext{6}{Departamento de Astrof{\'i}sica, Universidad de La Laguna, E-38206 La Laguna, Tenerife, Spain}
\altaffiltext{7}{IRFU, CEA, Universit\'e Paris-Saclay, F-91191 Gif-sur-Yvette, France}
\altaffiltext{8}{AIM, CEA, CNRS, Universit\'e Paris-Saclay, Universit\'e Paris Diderot, Sorbonne Paris Cit\'e, F-91191 Gif-sur-Yvette, France}
\altaffiltext{9}{Space Science Institute, 4750 Walnut Street Suite 205, Boulder, CO 80301, USA}

\begin{abstract}
Stars between two and three solar masses rotate rapidly on the main sequence, and the detection of slow core and surface rotation in the core-helium burning phase for these stars places strong constraints on their angular momentum transport and loss. 
From a detailed asteroseismic study of the mixed-dipole mode pattern in a carefully selected, representative sample of stars, we find that slow core rotation rates in the range reported by prior studies are a general phenomenon and not a selection effect. We show that the core rotation rates of these stars decline strongly with decreasing surface gravity during the core He-burning phase. We argue that this is a model-independent indication of significant rapid angular momentum transport between the cores and envelopes of these stars. We see a significant range in core rotation rates at all surface gravities, with little evidence for a convergence towards a uniform value.  We demonstrate using evolutionary models that measured surface rotation periods are a biased tracer of the true surface rotation distribution, and argue for using stellar models  
for interpreting the contrast between core and surface rotation rates. The core rotation rates we measure do not have a strong mass or metallicity dependence. 
We argue that the emerging data strongly favors a model where angular momentum transport is much more efficient during the core He burning phase than in the shell burning phases which precede and follow it.

\end{abstract}


\section{Introduction}
\setcounter{footnote}{0}

Real stars rotate.  Understanding the impact of rotation on the life cycle of stars, however, is one of the great outstanding problems in the theory of stellar evolution. Despite the maturity of the field, we currently lack a consensus model for the dominant internal angular momentum transport mechanism in stars. A major reason for this difficulty has been the lack of data on the internal rotation rates of stars. We can now use the tools of asteroseismology to infer core rotation rates in evolved giant stars, which is transforming our understanding of the field.

Intermediate-mass stars (M $\sim$ 2.0 - 3.0 M$_{\sun}$) provide important tests of the role of rotation in the structure and evolution of stars because they live in an important transitional regime. Like massive stars, they rotate rapidly on the main sequence \citep[\vsini\ up to 300 \kms,][]{ZorecRoyer2012} and have convective cores. However, after core hydrogen exhaustion, these stars quickly cross the Hertzsprung gap and become cool red giants. Like other cool giants, these stars are solar-like oscillators, and asteroseismology can be used to measure their bulk properties, like mass and radius, as well as finer details, like the core rotation rate \citep{Beck2012}. 

Intermediate-mass stars have a strongly mass dependent lifetime on the first ascent red giant branch and reach a mass dependent luminosity at the tip. Stars above about 2.25 M$_\sun$ then nondegenerately ignite their helium and contract down to the core-helium burning sequence. The mass range we consider overlaps with the empirically identified secondary clump \citep[2RC, 1.7-2.5 \msun,][] {Girardi1998}, and so we use that term to distinguish between the core-helium burning phase of these intermediate-mass stars and the core-helium burning phase of lower mass stars (RC) or the shell hydrogen burning first ascent red giant branch phase (RGB). We note that this use of terminology is consistent with recent publications on the seismic identification of evolutionary states \citep{Bedding2011, Mosser2015}
using the period spacing of mixed dipole modes \citep{Beck2011}.

On the main sequence, stars between two and three solar masses have a median rotation rate around 150 \kms, with a broad distribution 
\citep{ZorecRoyer2012}. Recent work has indicated that core rotation rates of these stars are correlated with and similar to the surface rotation rates {\citep{vanReeth2016, Kurtz2014, Ouazzani2018, Li2019}}, although these stars are not necessarily perfect rigid rotators \citep{Kurtz2014}.

We also have asteroseismic data for evolved secondary clump stars.  They have core rotation periods of around 100 days \citep[around 0.2 $\mu$Hz,][]{Mosser2012b, Deheuvels2015}. This is slower than the rotation rates of the cores of low-mass stars on the subgiant and first ascent giant branch \citep[around 10 days,][]{Mosser2012b, Deheuvels2012,Deheuvels2014, Gehan2018} and similar to the rotation rates of cores on the red clump \citep{Mosser2012b}. {While \cite{Beck2012} found that the cores of low-mass stars at the beginning of the red giant phase rotate 10 times faster than the surface,} \citet{Deheuvels2015} found that the cores of secondary clump stars rotate between one and three times faster than the envelopes. This is significantly closer to solid-body rotation than the subgiants studied in \citet{Deheuvels2014} and consistent with limiting case theoretical arguments that strong core-envelope coupling must be occurring in secondary clump stars \citep[][{see also \citealt{Aerts2019} for a more complete discussion of the available constraints on angular momentum evolution}]{TayarPinsonneault2013, Cantiello2014, denHartogh2019}. However, due to the size of previous samples and the methods by which they were selected, it has been difficult to determine whether they accurately represent the full underlying distribution of core rotation rates in the secondary clump.

We also have measured surface rotation rates for secondary clump stars from both the projected surface rotation velocity, \vsini, from spectroscopy \citep[e.g.][]{Tayar2015} and the rotationally modulated signature of star spots from space photometry \citep{Ceillier2017}. These measurements can be used to test whether angular momentum evolution models developed for the spin down of solar-type stars can be applied to the secondary clump. These models commonly use the \citet{Kawaler1988} magnetized-wind formulation to infer predicted torques, and assume the entire star rotates as a solid body. However, these recent measurements indicate that the surfaces of these stars rotate much more slowly than expected, suggesting either enhanced angular momentum loss compared to the Kawaler-calibrated models or substantial radial differential rotation \citep{ Tayar2015,Ceillier2017, TayarPinsonneault2018}. 
Work by \citet{TayarPinsonneault2018} demonstrated that enhanced angular momentum loss at the level of the Pinsonneault, Matt, and MacGregor (PMM) wind-loss law \citep{vanSadersPinsonneault2013} is required to match the measured rotation distributions. They also showed that some amount of radial differential rotation is required by the data, but were unable to determine whether that differential rotation was taking place in the core of the star or its envelope.

In this paper, we focus on measuring core rotation rates in a carefully selected sample of secondary clump stars in order to understand the core rotation distribution and any trends with physical parameters such as mass, metallicity, surface gravity, and surface rotation, which might prove useful in understanding the angular momentum transport, angular momentum loss, and resulting radial rotation profile in these stars. Based on the work from \citet{TayarPinsonneault2018} we can also robustly predict surface rotation periods as a function of mass, metallicity and surface gravity. We can therefore both use theoretically predicted surface rates and the available measurements to study the correlation between core and surface rotation rates in our sample.  We also have detailed spectroscopic information and can look for trends in core rotation as a function of the measured surface abundances.  In this way we can both test for correlations with metallicity and with the carbon to nitrogen ratio, which is a diagnostic of the first dredge-up. We also include a study of limiting case theoretical scenarios for internal angular momentum transport.

\section{Data}

\subsection{General Seismic and Spectroscopic Parameters} \label{Sec:globalparams}
There has been significant recent progress on both spectroscopic \citep{DR14} and asteroseismic \citep{Pinsonneault2018} data for targets in the \textit{Kepler} fields. However, our sample selection pre-dates those works, and  
as a result our sample selection was designed using related, but different, data sets.  We therefore begin by discussing the tools we used to define our candidates for analysis.

For sample selection we use data collected by the \textit{Kepler} mission \citep{Borucki2010} that has monitored more than 197,000 stars \citep{Mathur2017}. 
The lightcurves analyzed in this work were first calibrated with the KADACS \citep[Kepler Asteroseismic Data Analysis Calibration Software,][] {Garcia2011,Garcia2014} software to remove outliers, trends, instrumental effects, and concatenate the quarters. All the stars in our sample have a sampling of 29.40min and have been observed for around four years; lightcurves are available at the MAST website\footnote{https://archive.stsci.edu/prepds/kepseismic/}. 

We applied the A2Z pipeline \citep{Mathur2010} to a sample of evolved stars with \kepler\ data. For all targets we inferred two global seismic parameters: the mean large frequency separation ($\Delta \nu$,  the frequency difference between two modes of the same degree and consecutive orders) and the frequency of maximum {oscillation} power ($\nu_{\rm max}$). We choose this particular asteroseismic analysis pipeline because it has also analyzed stars with visible surface modulation that allows the measurement of a star spot rotation period \citep{Ceillier2017} but can complicate asteroseismic analysis.

Spectroscopic parameters for these stars, including temperature and metallicity, come from the APOGEE spectroscopic survey \citep{Majewski2017}. Specifically, we used data from Data Release 12 \citep[DR12,][]{Alam2015} of the Sloan Digital Sky Survey III \citep{Eisenstein2011} for our target selection process. Spectra were taken on the 2.5m Sloan Digital Sky Survey telescope \citep{Gunn2006} and analyzed with the ASPCAP pipeline \citep{Nidever2015,GarciaPerez2015}. The stellar parameters for DR12 were calibrated using both members of star clusters and asteroseismic data \citep{Meszaros2013}. 

Since these targets were selected, improved measurements for both the spectroscopic and asteroseismic parameters have been released. For the analysis here, we use the spectroscopic values from the Fourteenth Data Release \citep{DR14} of the Sloan Digital Sky Survey IV \citep{Blanton2017} which takes advantage both of an updated spectroscopic analysis pipeline and improvements in the calibration to fundamental measurements \citep{Holtzman2018}. Where possible, we also update our seismic parameters to those published in \citet{Pinsonneault2018}, which has compared multiple analysis techniques, corrected for known departures from the asteroseismic scaling relations, and placed the results on a fundamental mass scale using open clusters. For the 28 stars where masses and surface gravities are not available in \citet{Pinsonneault2018}, we use our A2Z pipeline data. \citet{Pinsonneault2018} defined normalizations for asteroseismic measurements that put all methods there, including A2Z, on a common basis, and we adopt those normalizations here (effective solar $\Delta\nu$ and $\nu_{\rm max}$ of 135.146 $\mu$Hz and 3091.44 $\mu$Hz respectively, as well as the A2Z specific correction factors). {We note that in some cases these updates cause stars to fall outside of our targeted mass range. For completeness, we still include and report the results for these stars, but we note that their exclusion would not substantially change the results of our analysis.}

\subsection{Sample Selection}

\begin{table*}[tbp]

\begin{minipage}{1.0\textwidth}
\caption{Properties of the stars studied here. Mass and surface gravity estimates come from \citet{Pinsonneault2018} or the A2Z pipeline with \citet{Pinsonneault2018} correction factors. Temperatures, metallicities and [C/N] come from APOGEE DR14 \citep{DR14}; periods come from \citet{Ceillier2017}. \DPi, coupling parameters (q), rotational splitting values, and core rotation periods are derived as discussed in Section \ref{Sec:periodspacing}. Theoretical surface rotation periods are derived using the models discussed in Section \ref{Sec:models}. Groups indicate why a star was selected: (1) in \citet{Deheuvels2015}, (2) spectroscopic selection, Section \ref{sssec:unbiasedsel}, (3) surface rotation period from \citet{Ceillier2017}, (4) asteroseismic sample, Section \ref{sssec:seismicsel}. Table 1 is published in its entirety in the machine-readable format; a portion is shown here for guidance regarding its form and content.}

\begin{tabular}{rrrrrrrrrrrrr}

\multicolumn{1}{c}{KIC} & \multicolumn{1}{c}{Mass} & \multicolumn{1}{c}{log(g)} & \multicolumn{1}{c}{T$_{\rm eff}$} & \multicolumn{1}{c}{[Fe/H]} & \multicolumn{1}{c}{[C/N]} & \multicolumn{1}{c}{P$_{\rm surf}$} & \multicolumn{1}{c}{Group} & \multicolumn{1}{c}{\DPi} &\multicolumn{1}{c}{q} & \multicolumn{1}{c}{splitting} & \multicolumn{1}{c}{P$_{\rm core}$} & \multicolumn{1}{c}{P$_{\rm surf, theory}$} \\ 
\multicolumn{1}{c}{} & \multicolumn{1}{c}{(\msun)} & \multicolumn{1}{c}{(dex)} & \multicolumn{1}{c}{(K)} & \multicolumn{1}{c}{(dex)} & \multicolumn{1}{c}{(dex)} & \multicolumn{1}{c}{(days)} & \multicolumn{1}{c}{} & \multicolumn{1}{c}{(s)} &\multicolumn{1}{c}{} & \multicolumn{1}{c}{($\mu$Hz)} & \multicolumn{1}{c}{(days)} & \multicolumn{1}{c}{(days)} \\
\hline \hline
3744681 & 2.315 & 2.71 & 4938.10 & 0.042 & -0.648 & -9999 & 1 & 285.9 &0.3 & 0.1 & 57.87 & 152.12 \\ 
4659821 & 2.23 & 2.933 & 5032.54 & 0.098 & -0.662 & -9999 & 1 & 213.4 &0.2 & 0.085 & 68.08 & 69.12 \\ 
5184199 & 2.165 & 2.905 & 5000.37 & 0.102 & -0.816 & -9999 & 1 & 239.2 &0.2 & 0.095 & 60.92 & 92.55 \\ 
7467630 & 2.765 & 2.78 & 4998.47 & 0.138 & -0.674 & -9999 & 1 & 293.3 &0.3 & 0.06 & 96.45 & 76.94 \\ 
7581399 & 2.894 & 2.841 & 5079.48 & 0.022 & -0.679 & 150.83 & {1,3} & 221.35&0.25 & 0.08 & 72.34 & -9999. \\ 
8962923 & 2.158 & 2.828 & 5051.95 & -0.056 & -0.532 & -9999 & 1 & 298.2&0.3 & 0.075 & 77.16 & 118.07 \\ 
9346602 & 2.141 & 2.658 & 4908.06 & -0.011 & -0.529 & -9999 & 1 & 242.8 &0.3 & 0.08 & 72.34 & 196.53 \\ 
8805663 & 2.325 & 2.605 & 4885.94 & 0.134 & -0.883 & -9999 & 2 & 325.2 &0.3 & 0.05 & 115.74 & 204.40 \\ 
8414518 & 2.314 & 2.665 & 4930.66 & 0.128 & -0.603 & -9999 & 2 & 267.45 &0.3 & 0.06 & 96.45 & 177.31 \\ 
8872979 & 2.576 & 2.801 & 5119.89 & 0.028 & -0.883 & -9999 & 2 & 297.5 &0.3 & 0.075 & 77.16 & 88.10 \\ 
9002805 & 2.087 & 2.735 & 5013.67 & 0.041 & -0.653 & -9999 & 2 & 243.7 &0.3 & 0.06 & 96.45 & 169.36 \\ 
\hline
\end{tabular}
\label{Table:data}
\end{minipage}
\end{table*}

\subsubsection{Comparison sample}\label{sssec:dehsel}
There are seven stars with published core rotation rates in this regime from \citet{Deheuvels2015}. In order to validate our analysis technique, which is somewhat simpler than what was used in that work, we add all of these stars to our sample and compare our results to theirs in Section \ref{coremeasurement}. 

\subsubsection{Unbiased sample}\label{sssec:unbiasedsel}

\begin{figure}
\begin{center}
\subfigure{\includegraphics[width=9cm, clip=true, trim=0.3in 0in 0in 0in]{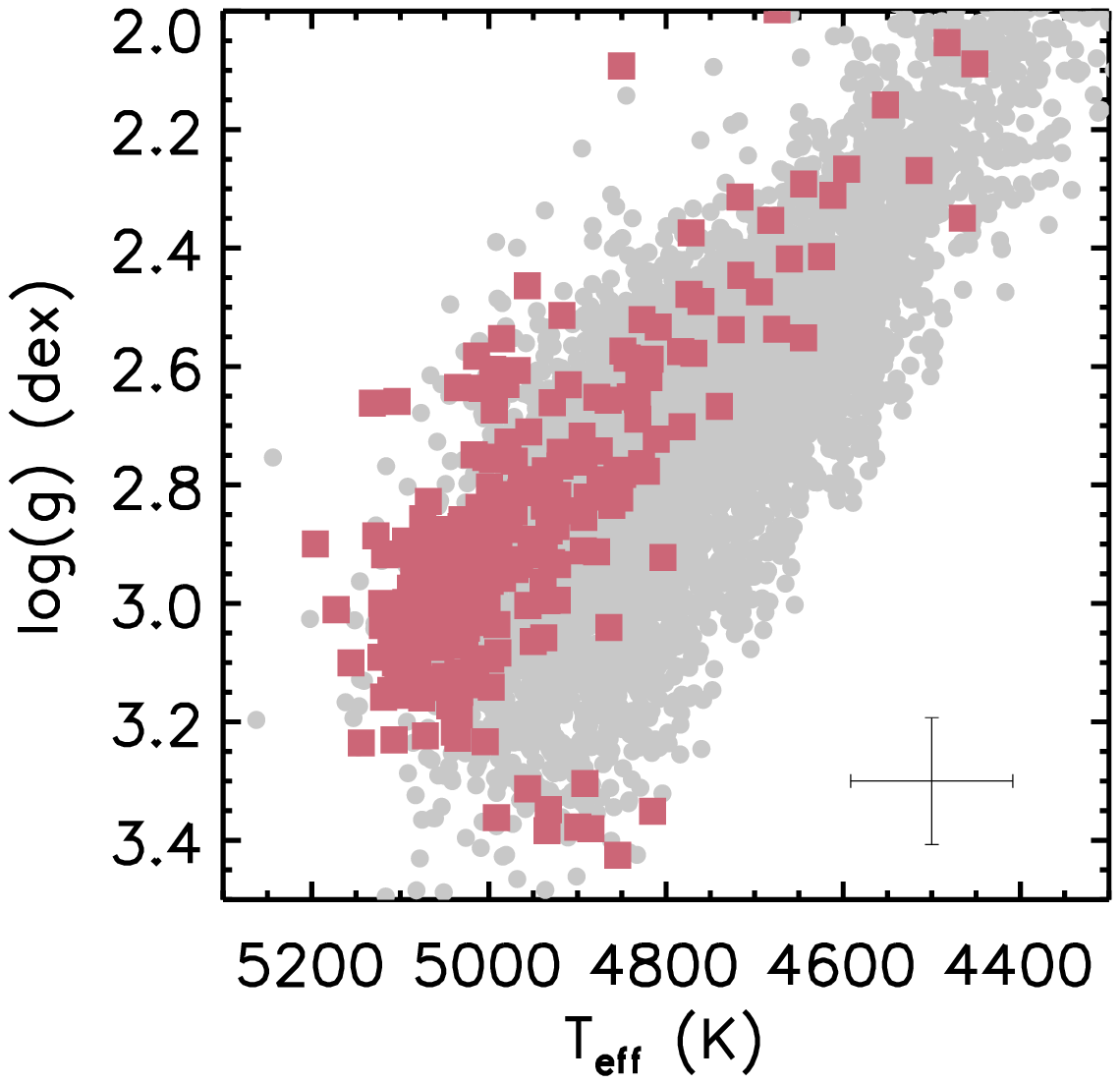}}
\subfigure{\includegraphics[width=9cm, clip=true, trim=0.3in 0in 0in 0in]{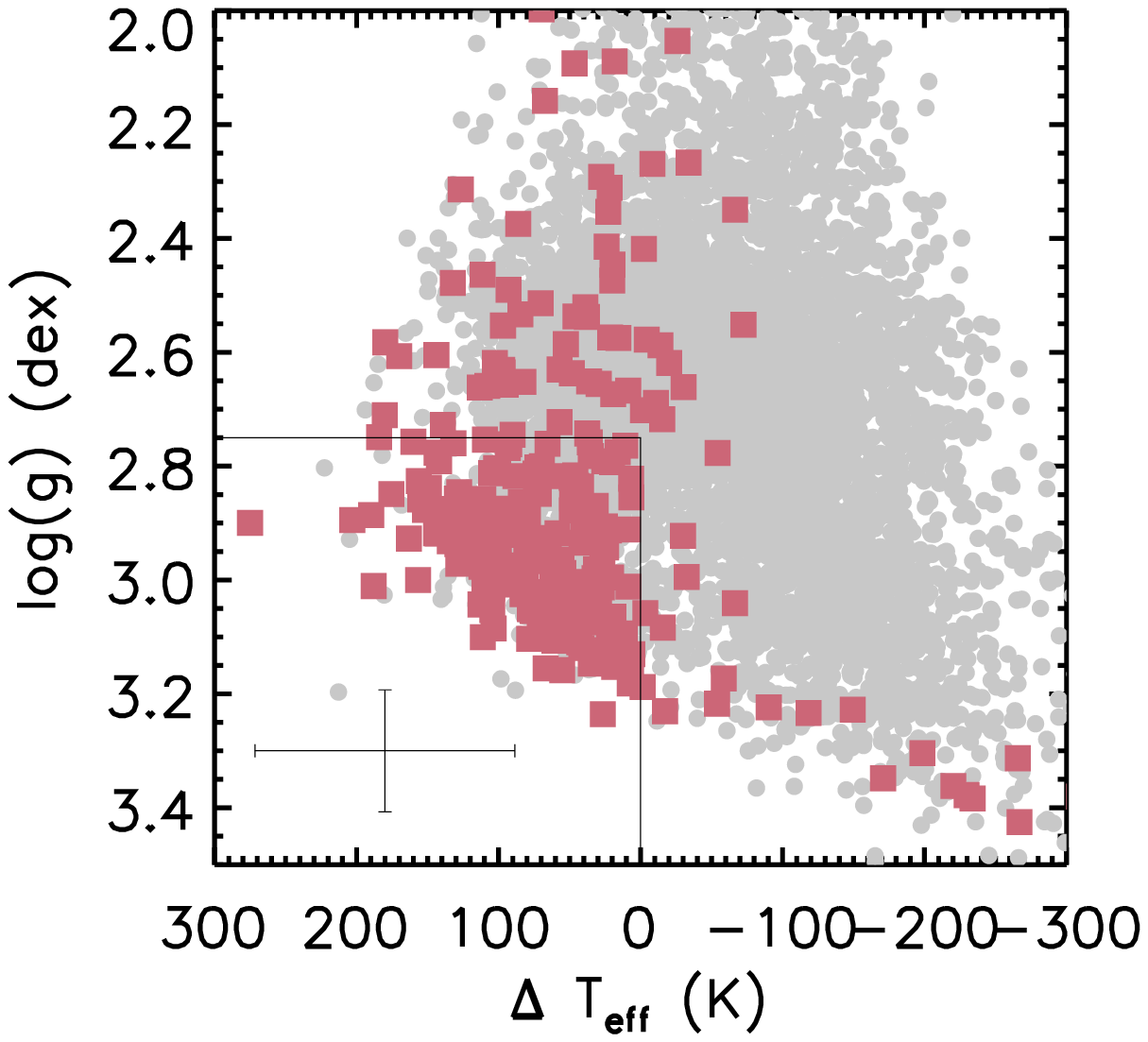}}
\caption{Top: The distribution of stars in temperature and gravity at the time of selection. The full sample is shown as grey circles, stars with preliminary masses above 2.3 \msun\ are highlighted as pink squares with characteristic error bars shown in the lower right corner. Bottom: Surface gravity versus offset from the predicted temperature dividing the clump and first-ascent-giant stars, which is surface gravity and metallicity dependent. We use the \citet{Bovy2014} formulation as discussed in Section \ref{sssec:unbiasedsel} to identify the region of parameter space where intermediate-mass stars are most common.}

\label{Fig:SpecSel}
\end{center}
\end{figure}

While prior samples of measured core rotation rates were generally selected to emphasize stars with strong, reliable detections, we are particularly interested in the statistical properties of the distribution of core rotation rates, which implies that we want an unbiased selection function. While we cannot control whether \kepler\ observed the stars, or the detectability of the modes required to measure core rotation, we can at least select a representative sample. We therefore chose a particular region of the sky, defined a sample spectroscopically, and looked at \kepler\ targets observed within that window. Specifically, 
we used two pointings of the APOGEE Data Release 12 sample in the \textit{Kepler} field. The stars in these fields represent all of the red giants in the region within a color and magnitude limited sample \citep{Zasowski2017}.  
We then defined, using the distribution of seismically measured masses and evolutionary states in \citet{Pinsonneault2014}, the region of spectroscopic gravity and temperature where massive, core-helium burning stars were most likely to be found (see Figure \ref{Fig:SpecSel}). We used the \citet{Bovy2014} red clump catalog separation between core-helium burning and shell hydrogen burning stars as a function of gravity and temperature as a starting point, and then optimized the slope and intercept to better separate seismically identified massive, core-helium burning stars, which tend to be higher metallicity, hotter, and higher gravity than low mass red clump stars. We used the relation: \[{\bf2.75} \leq {\rm log}\,g \leq 0.0018 {\rm\, dex \,K}^{-1} (T_{\rm eff}-T^{\rm ref}_{\rm eff}({\rm \,[Fe/H]}))+{\bf2.3}\]  where \[T^{\rm ref}_{\rm eff}({\rm [Fe/H]})={\bf -201.6} {\rm \,K \,dex}^{-1}{\rm [Fe/H]}+4607 {\rm K} \]
with bolded values indicating our alterations to the original \citet{Bovy2014} formula.
Analysis of the 669 stars meeting this criteria suggested that this cut was expected to yield a sample which was 28.2\% pure and 46.8\% complete for stars above 2.3 \msun.

\subsubsection{Surface Rotation Sample}\label{sssec:rotationsel}

Determining the surface rotation rate of a giant purely from an asteroseismic analysis is extremely difficult. This is because the non-radial oscillation modes which would be used to measure the surface rotation experience a significant contribution from the core rotation rate and a possible rotational gradient in the convective envelope, which has a stronger impact on the observed frequencies. As a result, inferring the surface rotation from asteroseismology alone requires detailed modeling and often gives only upper limits \citep{Deheuvels2014}. The ability to understand the radial rotation profile of a star benefits tremendously from either theoretical models or an independent measurement of the surface rotation from, e.g., the modulation of photometric data \citep[e.g.][]{Ceillier2017, Garcia2014b}, and such measurements are generally consistent with surface rotation rates measured with asteroseismology \citep[e.g.][]{Beck2018}. \citet{Ceillier2017} was able to measure surface rotation periods for about two percent of massive core-helium burning secondary clump stars, including sixty stars with APOGEE spectra. We added all sixty of these stars to our analysis sample. We caution, however, that the median measured surface rotation rate does not decrease with surface gravity as the stars expand, in contrast with the predictions of angular momentum conservation. This suggests that there are significant selection biases in the underlying sample of active stars with measurable rotation periods, likely due to the difficulty of measuring rotation periods substantially longer than a single \kepler\ quarter ($\sim$90 days). We consider the implications of these selection biases on the core-envelope contrasts inferred from these data in Section \ref{Sec:core-envelopecontrast}.

\subsubsection{Seismically Selected sample}\label{sssec:seismicsel}

For our main sample, which was selected using all available seismic and spectroscopic information, we test the dependence of core rotation on mass, gravity, and metallicity. 
To select the sample, we started with the sample of all \textit{Kepler} giants observed by the Sloan Digital Sky Survey III \citep[Data Release 12,][]{Alam2015}. We further selected only stars with metallicities and temperatures from the ASPCAP pipeline as well as seismic $\Delta\nu$ and $\nu_{max}$ measurements from the A2Z pipeline (R. Garc\'ia et al. in prep). We used the seismic scaling relations \citep{KjeldsenBedding1995} with the solar reference values appropriate to this pipeline to get a preliminary seismic mass, radius, and surface gravity for each star. For this initial selection, we did not apply corrections to the $\Delta\nu$ scaling relation, as they were not included in the \citet{Pinsonneault2014} analysis.  These corrections are in general small for these stars \citep{Pinsonneault2018}, so this does not significantly impact our selections. 
We identified two mass bins (2.2 to 2.4 M$_\sun$  and 2.6 to 2.8 M$_\sun$) and ordered the stars by metallicity. In the highest mass (youngest) bin, metallicities ranged from ([Fe/H]= -0.2 to +0.5). To cover this range, we selected two metallicity bins of interest, a low metallicity bin ([Fe/H]$<$-0.0) and a high metallicity bin ([Fe/H]$>$+0.2). Finally, we wanted to cover the evolution of stars in each mass and metallicity bin through the secondary clump. We therefore selected a star in each bin with a seismic surface gravity of approximately 3.0, 2.8, and 2.6 dex. While the limited number of stars in some bins required gravity offsets of several hundredths of a dex from our target, we were able to span a range of at least 0.3 dex in log(g) in each mass-metallicity bin. During the selection process, we emphasized stars with measured properties similar to active stars with measured surface rotation periods above in order to determine whether surface activity is correlated with core rotation rate.

\begin{figure}
\begin{center}
\subfigure{\includegraphics[width=9cm, clip=true, trim=0.3in 0in 0in 0in]{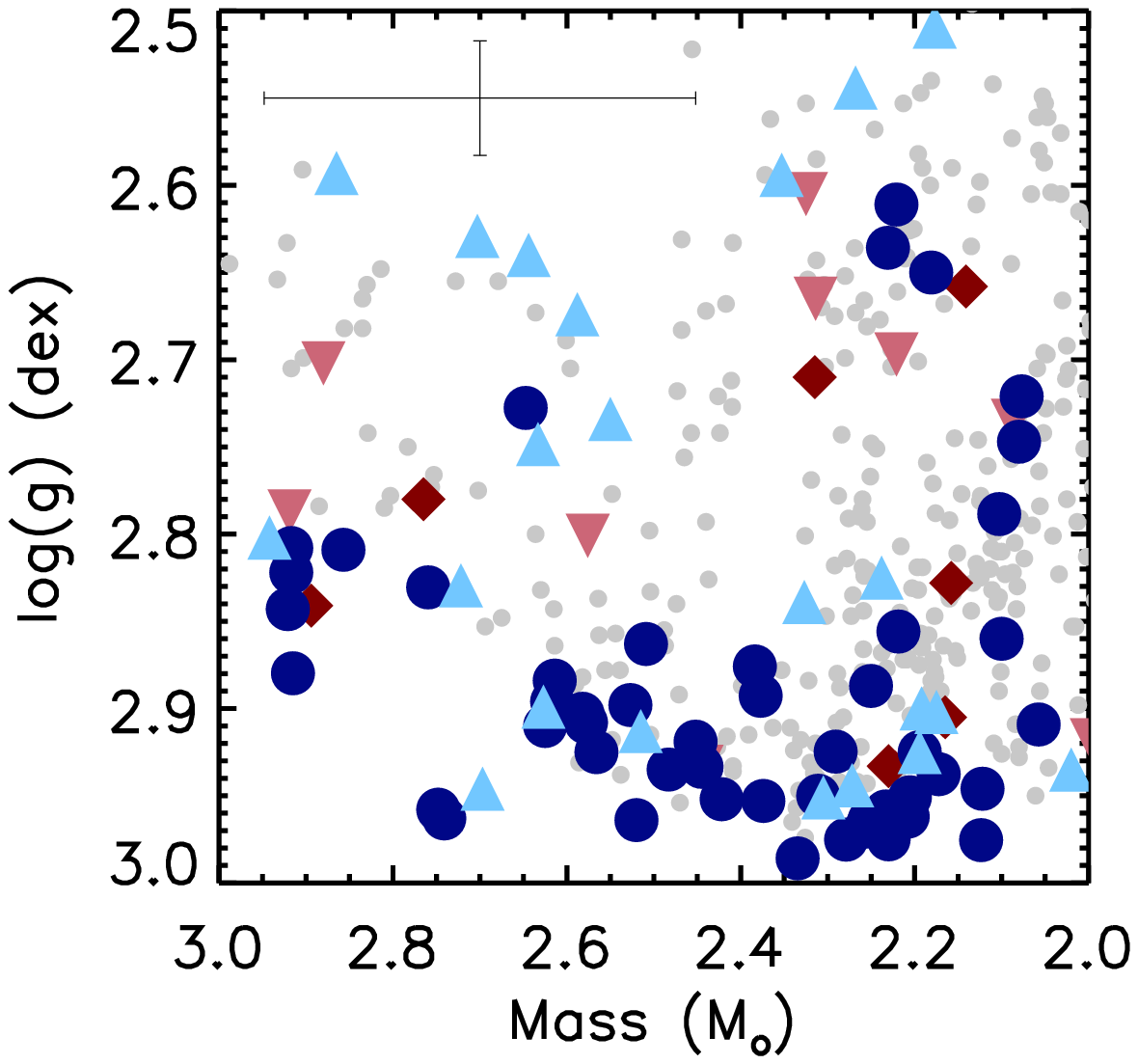}}
\subfigure{\includegraphics[width=9cm, clip=true, trim=0.3in 0in 0in 0in]{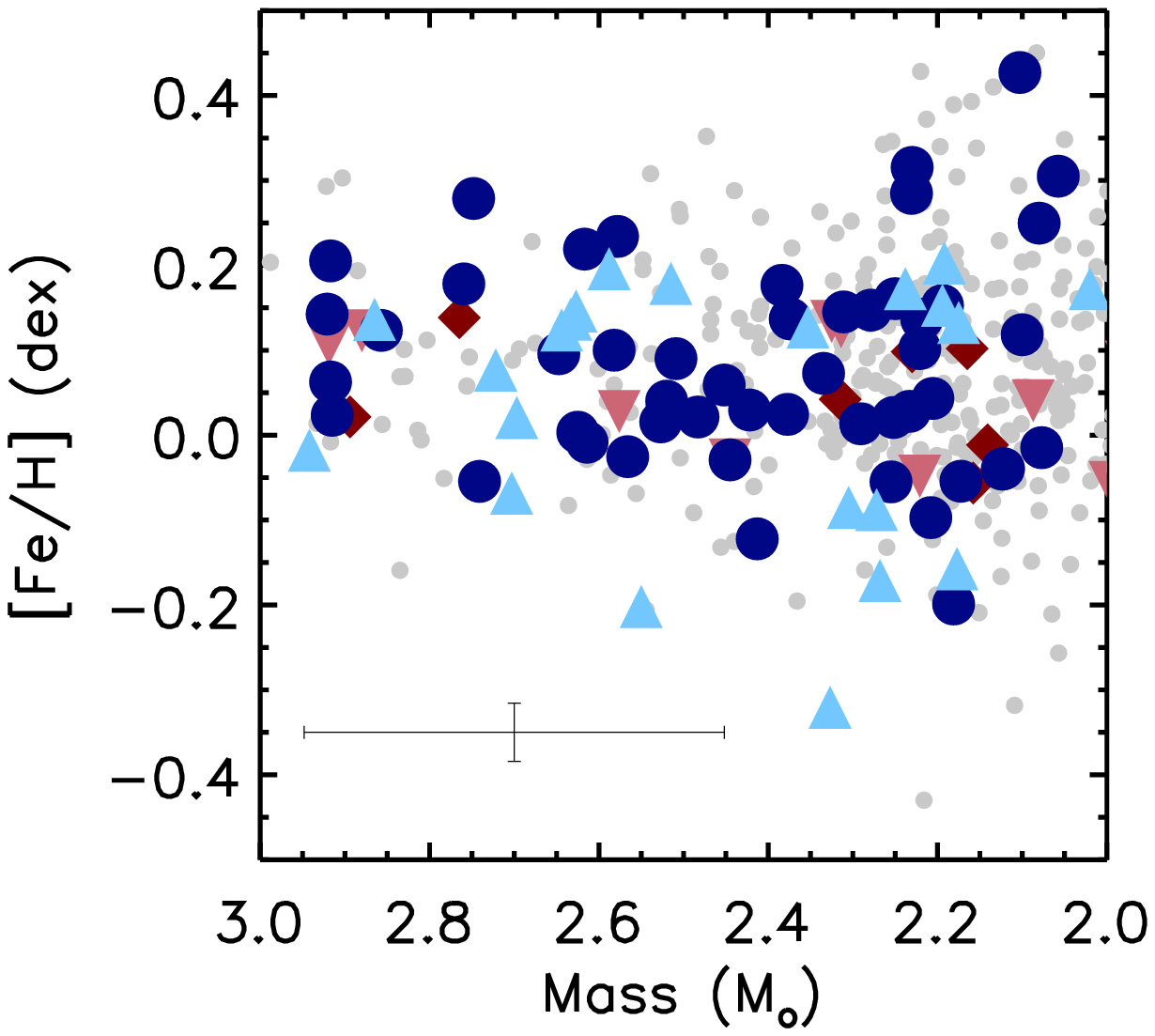}}
\caption{The stars in our sample (large points) compared to the distribution of stars in the full \citet{Pinsonneault2018} sample, shown as gray circles. Stars from the \citet{Deheuvels2015} sample are marked as red diamonds (Section \ref{sssec:dehsel}), stars from our spectroscopically selected sample are marked as pink downward triangles (Section \ref{sssec:unbiasedsel}), stars with measured surface rotation periods are marked as dark blue circles (Section \ref{sssec:rotationsel}) and stars selected using their seismic masses and gravities are shown as light blue upward triangles (Section \ref{sssec:seismicsel}). Characteristic error bars are shown to indicate the combined systematic and random uncertainties. }

\label{Fig:SeisSel}
\end{center}
\end{figure}

Evolved red giants develop deep surface convection zones, which incorporates nuclear processed material from the deep core into the surface layers.  This ``first dredge-up" changes the surface [C/N] ratio \citep{Iben1967}. There is a striking correlation between asteroseismic mass and the surface [C/N] ratio.  This has been used to infer masses, and thus ages, from spectra, using joint asteroseismic and spectroscopic data sets for training \citep{Martig2016,Ness2016}. However, there is a real intrinsic spread in [C/N] at fixed mass and metallicity \citep{Pinsonneault2018} for intermediate-mass stars, and it is interesting to explore whether it is correlated with stellar rotation.
We therefore identified three pairs and two trios of stars which had different [C/N] ratios ($\Delta$ [C/N] $>$0.2 dex) but otherwise were very similar ($\Delta$[Fe/H]$<$ 0.1 dex, $\Delta$[$\alpha$/Fe]$<$ 0.03 dex, $\Delta$Mass $<$ 0.13 M$_{\sun}$, and $\Delta$log(g) $<$ 0.05 dex). In all but one of our [C/N] groups, one of the stars was already included in one of the previous samples. This produced a final sample of 21 seismically selected stars, in addition to the 60 active stars, 20 giants in our spectroscopically selected sample, and seven stars with previously measured core rotation periods we use for calibration. All these stars and their properties are listed in Table \ref{Table:data}{, we show their distribution in mass, metallicity and surface gravity compared to the full \citet{Pinsonneault2018} sample in Figure \ref{Fig:SeisSel}, and their properties are summarized in Table \ref{Table:summary}.}

\begin{table}[htbp]
\caption{{Summary information for each of our samples. For stars that fit the criteria of multiple samples, we included them in only the first group to avoid double counting. Core period measurements are described in Section \ref{coremeasurement}, surface periods come from \citet{Ceillier2017}, and modeled periods are estimated from the grid described in Section \ref{Sec:models}.  Stars which fell outside the model grid in mass, metallicity, or mass dependent surface gravity do not have estimated model periods. The surface periods of the stars in the Seismic sample are marked with an asterisk because they were flagged as unreliable by \citet{Ceillier2017}.}}
\begin{tabular}{lrrrrr}

Name & \multicolumn{1}{l}{Group} & \multicolumn{1}{l}{Total } & \multicolumn{1}{l}{P$_{\rm core}$} & \multicolumn{1}{l}{P$_{\rm surf}$} & \multicolumn{1}{l}{Modeled} \\ \hline \hline
Comparison & 1 & 7 & 7 & 1 & 6 \\ 
Unbiased & 2 & 20 & 16 & 1 & 7 \\ 
Rotation & 3 & 58 & 32 & 58 & 35 \\ 
Seismic & 4 & 21 & 17 & 2* & 17 \\ \hline
Total & \multicolumn{1}{l}{} & 106 & 72 & 60 & 65 \\ \hline
\end{tabular}
\label{Table:summary}
\end{table}

\section{Mixed Mode Analysis}

\subsection{Determination of the evolutionary state \label{sec:evolState}}

To determine the evolutionary state of the stars, we
applied established seismic methodologies exploiting the period spacing of mixed dipole modes \citep[see e.g.][]{Bedding2011, Mosser2015}. 
For an initial guess we used
an autocorrelation analysis (AC) to determine the evolutionary stage for our sample by deriving the observed period spacing of the power spectral density (hereafter also PSD), which was converted from  frequency space into an equally spaced period spectrum. To improve the signal of the observed period spacing, we prewhitened the pure pressure radial modes and the pressure dominated quadrupole modes in the four central radial orders of the power excess. We fit Lorentzian profiles to both modes, and divided the PSD {by the fit.} 
This allows us to use the dipole mixed mode spectrum in {five} consecutive radial orders, without blurring the autocorrelation from the presence of the small separation between these modes or introducing aliases from the period gaps that occur if this frequency range is cut out. Because the highest peak might be a short periodic artifact, we search for the most likely peak in the autocorrelation by identifying the minimum of the second derivative of the autocorrelation response. Finally, the result of the pipeline is validated by visual inspection and comparison with the value ranges of \citet{Bedding2011} and \citet{Mosser2011a}.

\subsection{True Period spacing} \label{Sec:periodspacing}
After identifying and excluding any RGB stars present in the unbiased sample and confirming the 2RC status of the remaining stars, we performed a more elaborate analysis to find the precise parameters of the mixed mode spectrum and determine the core rotation period. 
From the asymptotic expansion based on work by \citet{Shibahashi1979} \citep[see also][] {Unno1989}, \citet{Mosser2012b} showed that for a star with known $\Delta\nu$ and $\nu_{\rm max}$ (our case, see Section\,\ref{Sec:globalparams}), the pattern of individual mixed dipole modes can be described through the use of the constant value of the period spacing, \DPi, which describes the actual but unobservable constant period spacing of dipole modes, and the corresponding coupling parameter $q$. Numerous techniques have been developed to determine the underlying value of  
\DPi and the corresponding coupling factor q for a given star \citep{Stello2012,Mosser2012b,Mosser2015}, 
and large samples of such measurements have been published \citep{Mosser2015,Vrard2016}.
Here we perform a grid search for the best parameter combination of  \DPi and $q$ for a given star. This approach was extensively explored by \citet{Buysschaert2016} but is very computational expensive. Therefore, the simplified version implemented by \citet{Beck2018} is used, based on the central radial mode, the large frequency separation, and a semi-automatically compiled list of frequencies of mixed modes ($|m|$=0).
The initial guess for the center of the \DPi-range is taken from the AC response of the observed period spacing. After running a coarse, wide grid covering about 200 seconds, a second grid search was performed around the minimum found in the first grid. The second grid was run with increments of 0.1\,seconds and 0.05 for \DPi and $q$, respectively, to avoid falling into a local minimum \citep[see][]{Buysschaert2016}.
The resulting values of \DPi and q are given in Table\,\ref{Table:data}.

To test the robustness and reliability of this approach, we used this method blindly on the 7 stars analyzed by \citet{Deheuvels2015}.
In all but one star, a nearly perfect agreement between the \DPi values of \citet{Deheuvels2015} and our analysis is found, and the reason for this single discrepancy is the different formalism used to fit the mixed mode pattern. Specifically, \citet{Deheuvels2015} used the revised mathematical description of the mixed mode pattern \citep[see e.g.][]{Mosser2014}, which involves 4 parameters instead of the original two parameters. However, visual inspection of the solution for the dipole forest shows good agreement between the two results, and we show in Figure \ref{Fig:omgcomp} that the inferred rotation rates are similar for the two methods, even with the discrepant assumptions for \DPi.

\subsection{Determination of the core-rotation rate} \label{coremeasurement}

Once the underlying parameters \DPi and $q$ were found for a given star, we searched for a g-dominated dipole mixed mode with a clear signature of rotational splitting. To test this estimate of the rotational splitting, we forced that value for the rotational splitting on all modes. This requires an assumption about the rotational gradient between the surface and the core, which, for simplicity, we assumed was a factor of two, corresponding to a constant value of the rotational splitting for all modes, regardless of the mode trapping. Because we used g-dominated modes to infer the splitting, we do not expect this assumption to substantially impact the inferred core rotation rates.
{If the found rotational splitting value of the g-dominated modes presented a convincing fit to the full range of dipole modes, we adopted the derived value as the rotational splitting of the core $\delta f_c$.} 
The resulting values of $\delta f_c$ are given in Table\,\ref{Table:data}.

For g-dominated modes, the rotational kernels are known to be dominated by the properties of the core \citep[e.g.][]{Beck2012,Beck2014, Goupil2013, Deheuvels2014, DiMauro2016, Triana2017}. As the Ledoux-constant \citep{Ledoux1951} is well approximated for these g-dominated mixed modes by 0.5, twice the value of the rotational splitting $\delta f_c$ is a very good proxy of the core rotation rate, and this is the value we report as our core rotation rate. 
We note that we did not extract the rotational splitting of pressure-dominated modes as their pattern is complicated by broad peaks originating from short mode lifetimes and the overall small level of rotational splitting.

\begin{figure}[!htb]
\begin{center}
\includegraphics[width=9cm,clip=true, trim=1in 0in 0in 0in]{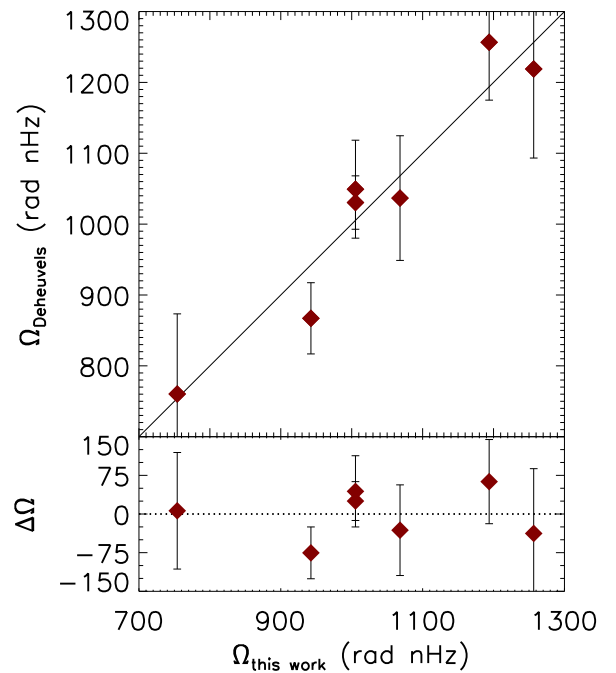}
\caption{Comparison between the core rotation rates computed in this work, and those published in \citet{Deheuvels2015}. We find that the agreement between the values (4\%) is better than the range of the published error bars (8\%). }
\label{Fig:omgcomp}
\end{center}
\end{figure}

\begin{figure}
\begin{center}
\includegraphics[width=9cm, clip=true, trim=.3in 0.1in 0in 0.3in]{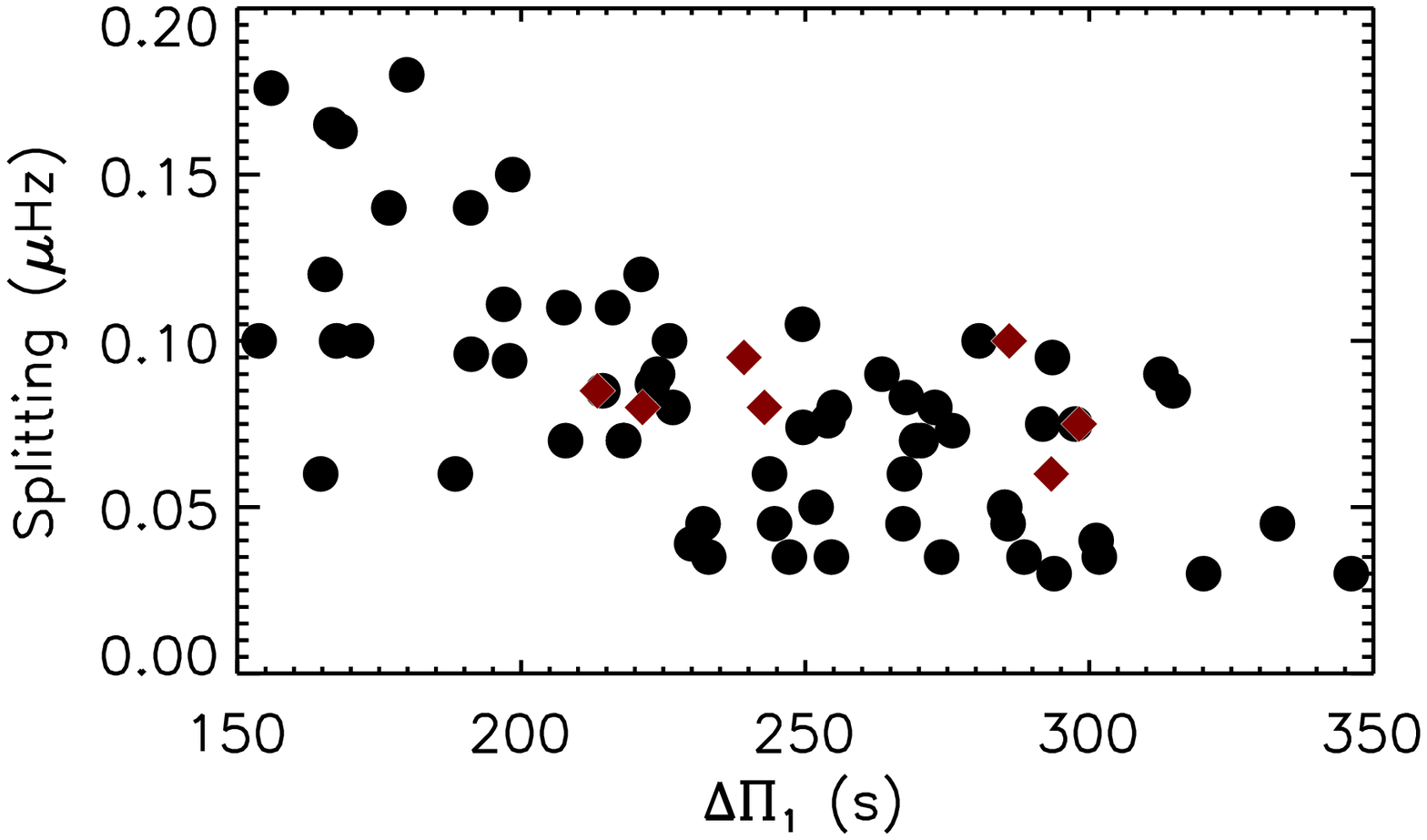}
\caption{The rotational splittings measured in this work as a function of \DPi. Red diamonds represent the values computed in this work for the stars published in \citet{Deheuvels2015}. There is a clear trend in splitting with \DPi, which was not visible in the \citet{Deheuvels2015} analysis because of the limited parameter space covered by that sample, and we discuss in later sections the constraints this trend places on models of angular momentum evolution. Errors are similar to the point size.}
\label{Fig:splitting}
\end{center}
\end{figure}

We compared the core rotation rates derived by this procedure with those determined by \cite{Deheuvels2015} (see Figure\,\ref{Fig:omgcomp}), using a somewhat different procedure, and we find agreement within 4\%. This is well within the formal uncertainty (8\%) of the sophisticated frequency analysis of \citet{Deheuvels2015} and we are confident that our method produces accurate and robust results. 
We show in Figure \ref{Fig:splitting} the resulting rotational splittings as a function of \DPi for the stars in our sample, as well as the \citet{Deheuvels2015} stars. Additionally, we choose to conservatively adopt the published error from \citet{Deheuvels2015}, eight percent, for the uncertainty on our core rotation measurements. Our wider range of \DPi allows us to see a trend in core rotational splitting with \DPi that was not visible using only the stars from \citet{Deheuvels2015} because of the limited parameter space covered by that sample.

\section{Models} \label{Sec:models}

In order to determine whether the trends in our data provide insight into the physics of rotational evolution, we compare the measured rotation rates to models of stellar evolution. For this purpose, we use a slightly expanded version of the grid of stellar models developed by \citet{TayarPinsonneault2018} to analyze the surface rotation rates of stars in this same mass range and phase of evolution. 

\subsection{Standard Model Physics}
As discussed in \citet{TayarPinsonneault2018}, these models are generated using the Yale Rotating Evolution Code \citep[YREC,][]{Pinsonneault1989, vanSadersPinsonneault2012} and cover the range from 2.0\msun\ to 3.0\msun\ in steps of 0.2\msun. Models are run using a \citet{GrevesseSauval1998} mixture, with atmosphere boundary conditions from \citet{Kurucz1997}. Core overshooting of 0.16 pressure scale heights has been used in order to reproduce the observed locus of the secondary clump \citep[see][]{TayarPinsonneault2018}. We have added to this grid models at two additional metallicities: [Fe/H]$=-0.3$ and [Fe/H]$=+0.3$, in addition to the solar metallicity models already included in \citet{TayarPinsonneault2018}. For these models at different metallicities, we have taken the calibration of the mixing length and helium abundance as a function of metallicity from \citet{Tayar2017}.  We also use the updated calculation of the convective overturn timescale discussed in \citet{Somers2017}.
We summarize the physical ingredients of the model and the extent of the model grid in Table \ref{Table:physics}.

\begin{table}[htbp]
\caption{Summary of the input physics used in our models. }
\begin{tabularx} {.48\textwidth}{>{\raggedright\arraybackslash}X|>{\raggedright\arraybackslash}X} 
\hline\hline
Parameter & YREC \\ \hline
Atmosphere & \citet{Kurucz1997}\\ 
$\alpha$-enhancement & No \\ 
Convective Overshoot & 0.16H$_{\rm p}$  \\ 
Diffusion & No  \\ 
Equation of State & OPAL+SCV \\ 
High-Temperature Opacities& OPAL \\ 
Low-Temperature Opacities & \citet{Ferguson2005}  \\ 
Mixing Length & \citet{Tayar2017}  \\ 
Mixture and Solar Z/X & \citet{GrevesseSauval1998} \\ 
Nuclear Reaction Rates & \citet{Adelberger2011} \\ 
Weak Screening & \citet{Salpeter1954}  \\ 
Solar X & 0.709306  \\ 
Solar Y & 0.272683  \\ 
Solar Z & 0.0179471  \\ 
Mass Range & 2.0 M$_\sun$ to 3.0 M$_\sun$ \\
Metallicities & [Fe/H]$=-0.3, 0.0, +0.3$\\\hline
\end{tabularx}
\label{Table:physics}
\end{table}

\subsection{Rotation Physics}

Analysis of the surface rotation rates of secondary clump stars have indicated that the surfaces of these stars are rotating much more slowly than standard models would predict \citep{Tayar2015, Ceillier2017,TayarPinsonneault2018}. \citet{TayarPinsonneault2018} showed that these slow rotation rates, in combination with the measured core rotation rates of \citet{Mosser2012b}, placed strong constraints on the allowable physics of angular momentum transport and loss in these stars. Specifically, the measured rotation rates require enhanced angular momentum loss relative to traditional wind laws, based on the \citet{Kawaler1988} framework, which were only weakly dependent on mass and radius.  More modern wind laws, in the \citet{Matt2012} framework, predict a stronger dependence on the underlying stellar parameters.  In this work, we employ the Pinsonneault, Matt \& MacGregor (PMM) wind law, as described in \citet{vanSadersPinsonneault2013}, 
which has an explicit dependence on the radius, luminosity, and convective overturn timescale and predicts 
enhanced spin down on the giant branch. This model has been successful in reproducing the observed rotation distributions of stars on the secondary clump \citep{TayarPinsonneault2018}, although matching previously observed distributions of core rotation periods seems to require some amount of radial differential rotation in the star, which could be happening either in the radiative interior or in the surface convection zone (or both). 

Previous works \citep[e.g.][]{TayarPinsonneault2013, Cantiello2014, denHartogh2019} have shown that the cores of these stars can not have been totally decoupled from their envelopes for their entire post-main-sequence evolution, as that would produce higher core rotation rates ($\Omega\sim10^{-2} \mu$Hz) and larger core-envelope contrasts ($\frac{\Omega_{core}}{\Omega_{surface}}\sim10^5$) than are observed. We therefore do not run separate calculations with totally decoupled core rotation here. 

Following the work of \citet{TayarPinsonneault2018}, we construct two classes of models whose predicted surface rotation distributions are consistent with the measured surface rotation distributions of intermediate-mass core-helium burning stars.

The simplest case is a star where both the core and surface rotate rigidly. This also serves as a lower limit on the range of allowed core rotation rates. 
Since the star is forced to rotate rigidly, this model also yields a slowly rotating core ($\Omega\sim10^{-7} \mu$Hz) and no difference between the core and envelope rotation rates ($\frac{\Omega_{core}}{\Omega_{surface}}=1$). One slight perturbation on this case is the possibility of a rigidly rotating envelope, but a differentially rotating core. Because the moment of inertia of the core is very small relative to the envelope, surface rotation period predictions for models with solid body rotation in the convection zone are insensitive to the treatment of internal angular momentum transport in the core, and so such models are still consistent with the available surface rotation constraints. In a few relevant plots, we add arrows to indicate the direction that differential rotation in the core would shift our predictions.

For our second case, 
we allow the differential rotation to take place in the surface convection zone, with a rotation profile that goes as $R^{-1}$. This model was shown in \citet{TayarPinsonneault2018} to be roughly consistent with the distribution of surface rotation rates, and we use it here as an upper limit on the predicted range of core rotation rates if the differential rotation is located in the surface convection zone. 
In this model, we fix the rotation of the core to the rotation rate at the base of the envelope, preventing a large shear from developing there. This model predicts a core that rotates substantially slower than the decoupled case, but faster than the surface of the star ($\Omega\sim10^{-6} \mu$Hz,$\frac{\Omega_{core}}{\Omega_{surface}}\sim3$).

\section{Analysis}
\subsection{Spectroscopically Selected Sample}
\begin{figure}[!htb]
\begin{center}
\includegraphics[width=9cm,clip=true, trim=0in 0in 0in 0in]{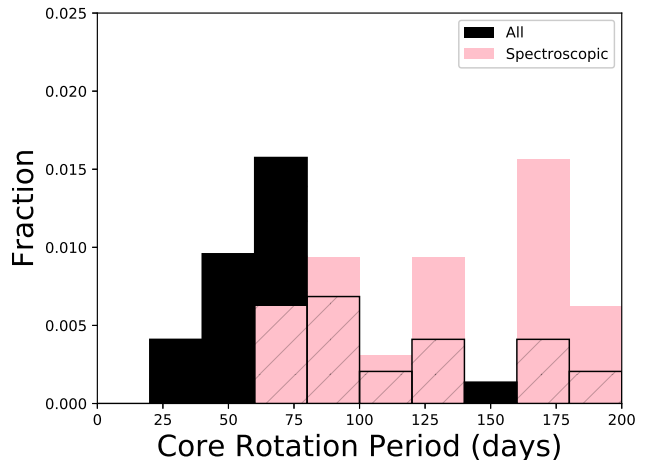}
\caption{Comparison between the core rotation rate distribution measured for the unbiased, spectroscopically selected sample and the full sample of stars analyzed in this work.} 
\label{Fig:seisspechist}
\end{center}
\end{figure}

With our spectroscopically selected sample, we can explore the true distribution of core rotation rates, and identify whether our seismically selected sample was biased \citep[e.g. against rapidly rotating stars, see][] {Tayar2015}. In the spectroscopically selected sample, where we expected 28\% of the stars to be above 2.3 \msun\ in the secondary clump, we ended up identifying 6/20 such stars (30\%), consistent with this prediction. We show in Figure \ref{Fig:seisspechist} that the distribution of core rotation rates for our spectroscopically selected sample is similar to our full sample, except that it lacks the fastest rotating stars. At first glance, this seems counter-intuitive, since larger rotational splittings and therefore shorter rotation periods are easier to measure. However, we note that the fastest rotators in the sample are all at very high gravity and often selected for their measurable surface rotation period or to fill out the highest gravity bins in our sample selected with seismic information. This suggests that such stars are over-represented in our sample, and we should be careful in future sections with any conclusions drawn solely from stars with surface gravities above about 2.9 or periods shorter than about 60 days. In particular, we acknowledge that given the size of our sample, some stars have almost certainly interacted with a companion on the upper giant branch, and these high gravity, rapidly rotating, active stars are consistent with the properties we expect for such interaction products, and thus should be treated carefully. Other than this fast rotating excess, however, our unbiased spectroscopically selected sample seems to match our full distribution of core rotation periods, and so we combine all four samples of stars for the following analysis.

\begin{figure}
\begin{center}
\subfigure{\includegraphics[width=8.5cm, clip=true, trim=0.3in 0.2in 0in 0.3in]{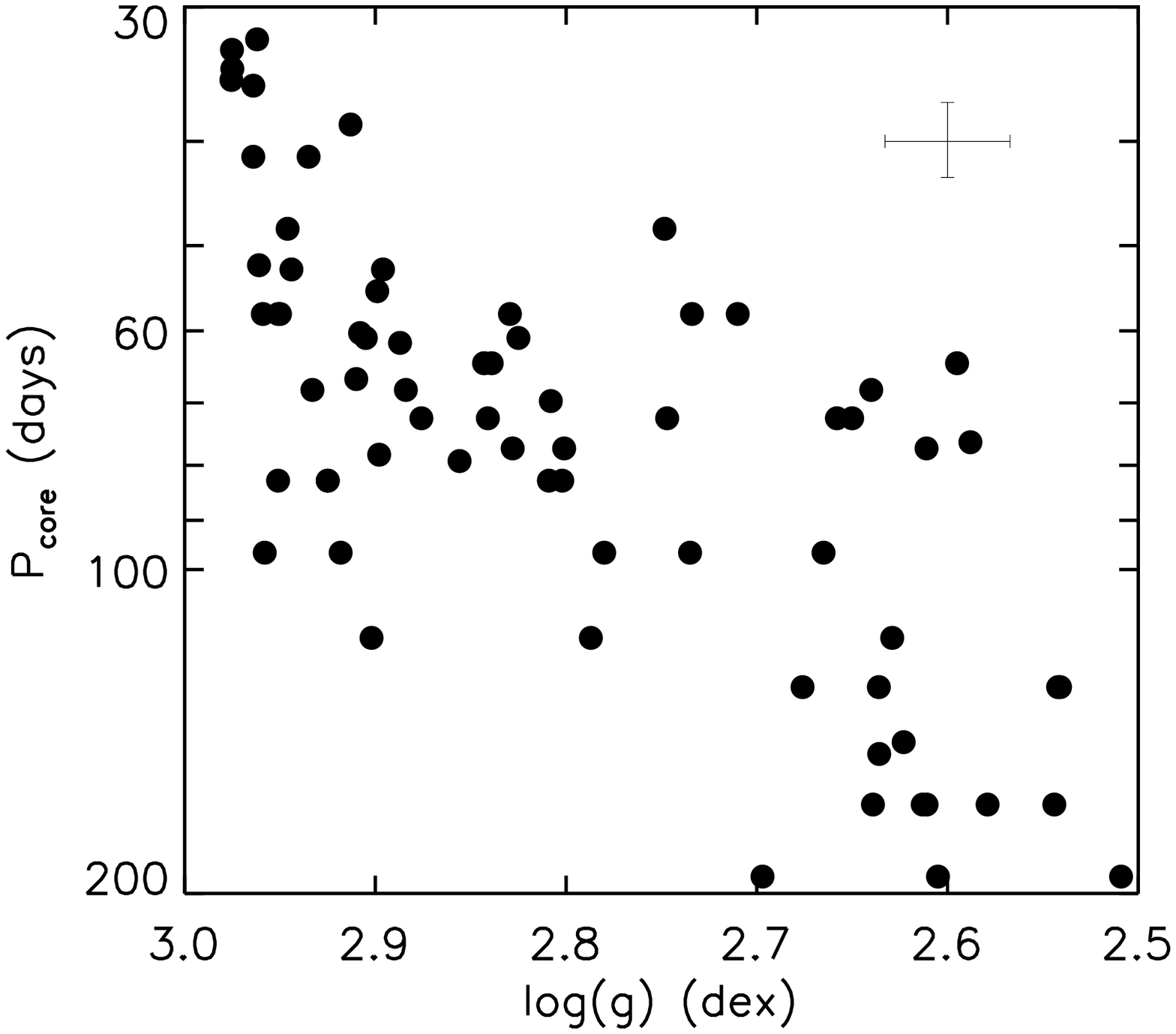}}
\subfigure{\includegraphics[width=8.5cm, clip=true, trim=0.3in 0.2in 0in 0.3in]{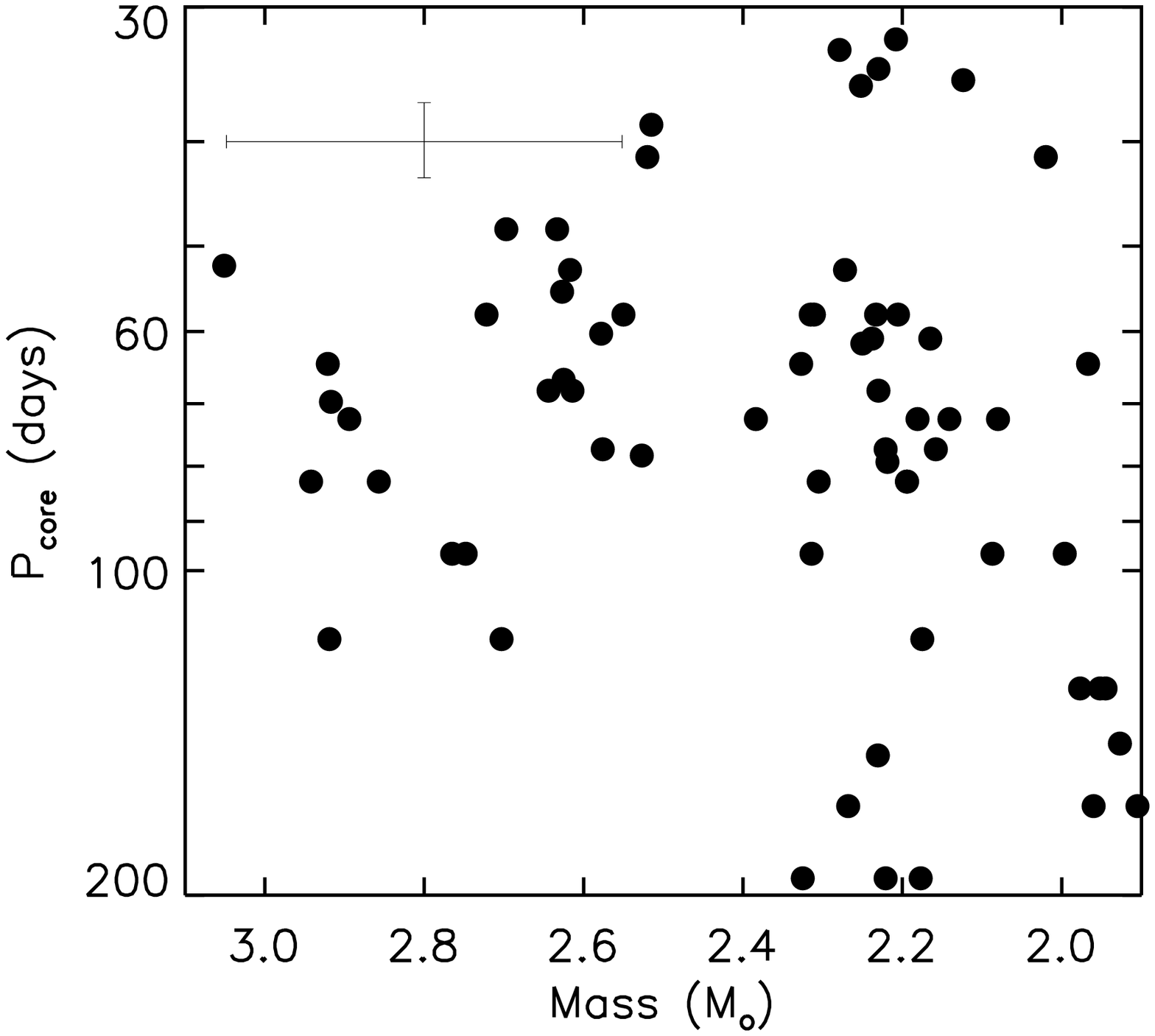}}
\subfigure{\includegraphics[width=8.5cm, clip=true, trim=0.3in 0.2in 0in 0.3in]{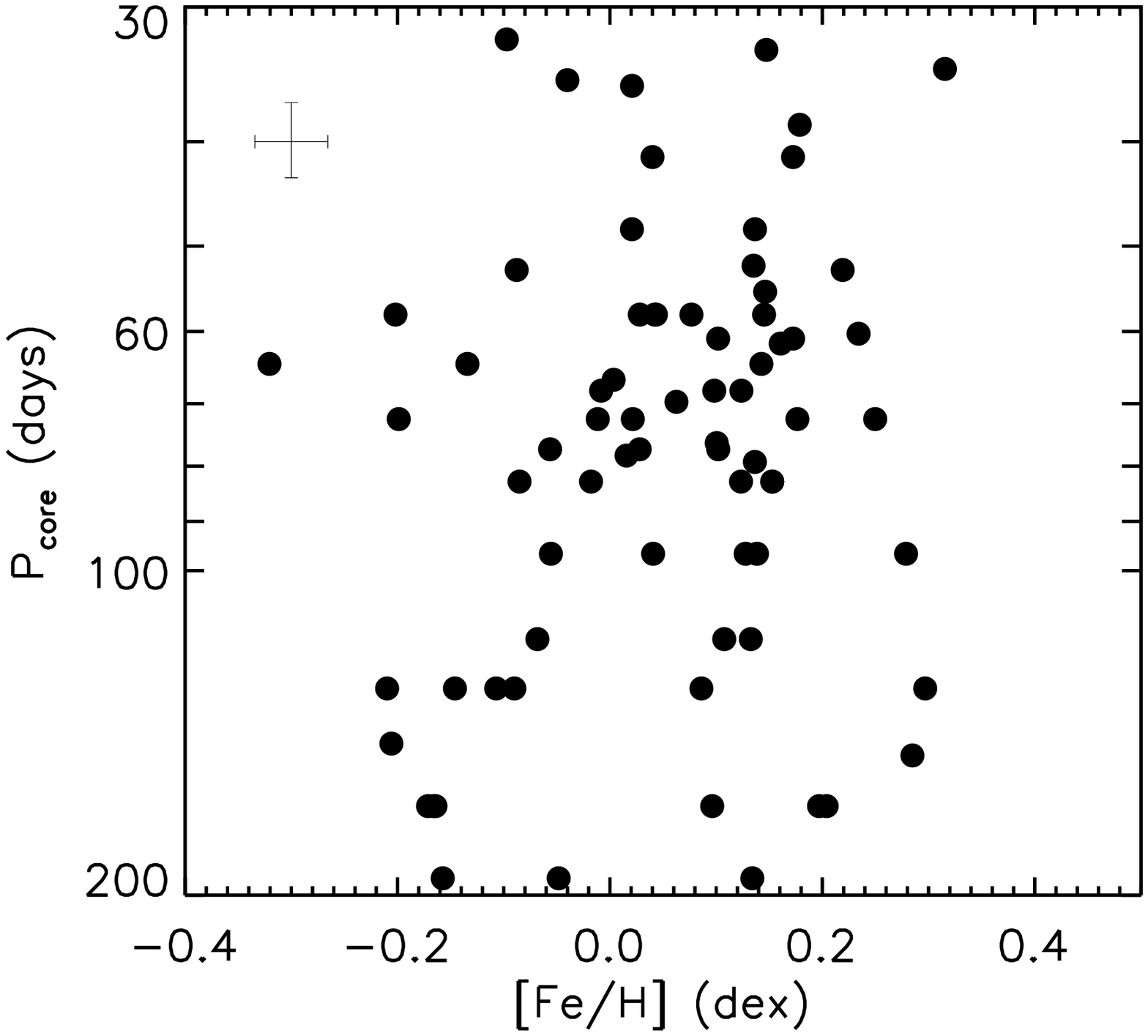}}
\caption{Comparison between the core rotation period and the surface gravity (top), mass(middle) and metallicity (bottom). We see a strong correlation with surface gravity and no significant correlation with either mass or metallicity. 
}
\label{Fig:Prots}
\end{center}
\end{figure}

\subsection{Core Rotation }
We show in Figure \ref{Fig:Prots} the core rotation rates we measured as a function of gravity (top), mass (middle) and metallicity (bottom). It is clear that the core rotation rate is correlated most strongly with surface gravity. Specifically, the cores of these stars are slowing down as the envelopes expand, providing the first model-independent evidence of strong core-envelope coupling on the secondary clump. We do not see any strong trends in the core rotation rates with mass or metallicity.

\subsection{Mixing and Magnetism}

\begin{figure}
\begin{center}
\subfigure{\includegraphics[width=8.cm, clip=true, trim=1.5in 0.0in 0.4in 0.5in]{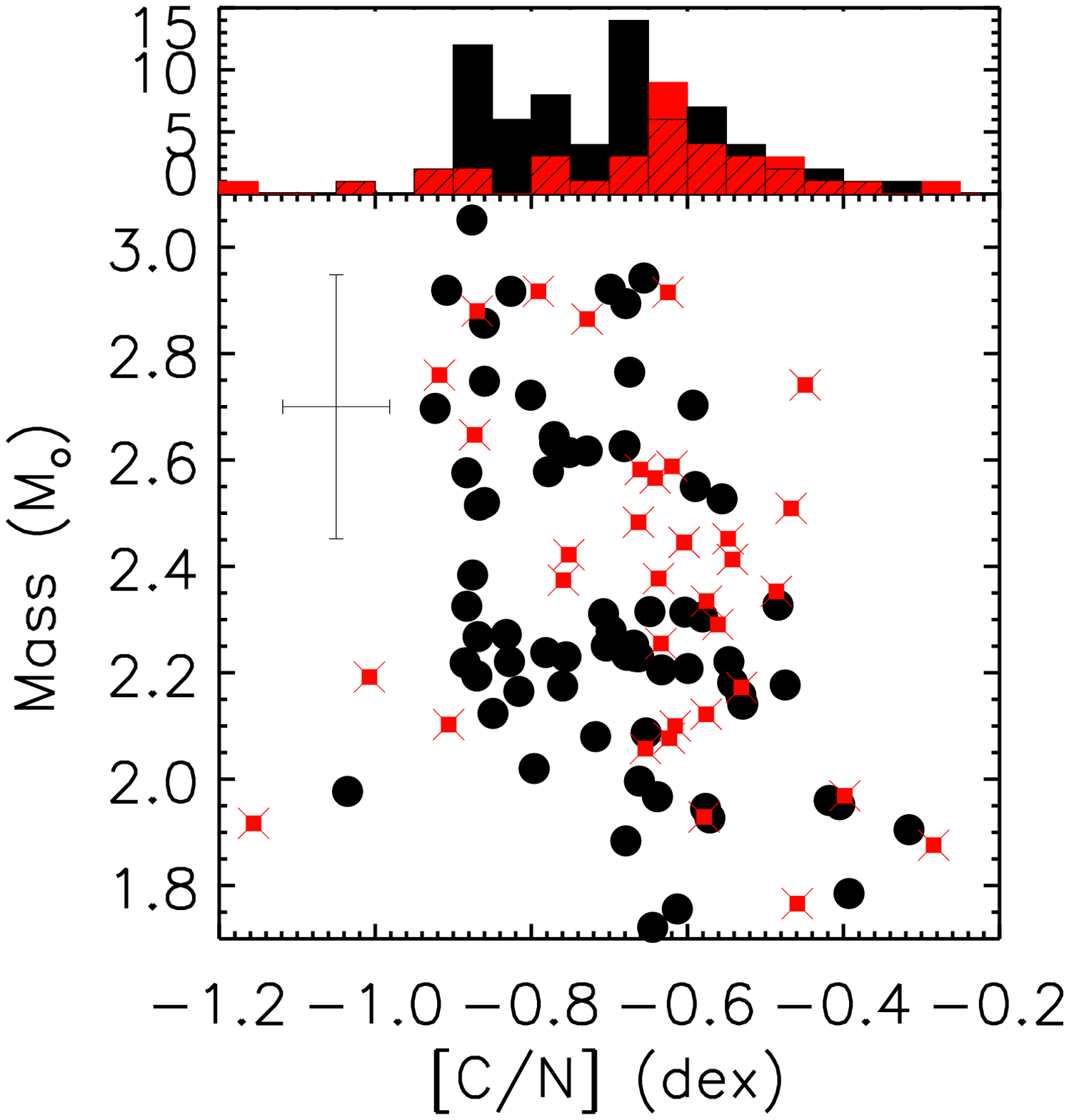}}
\subfigure{\includegraphics[width=8.cm, clip=true, trim=0.3in 0.2in 0.2in 0.5in]{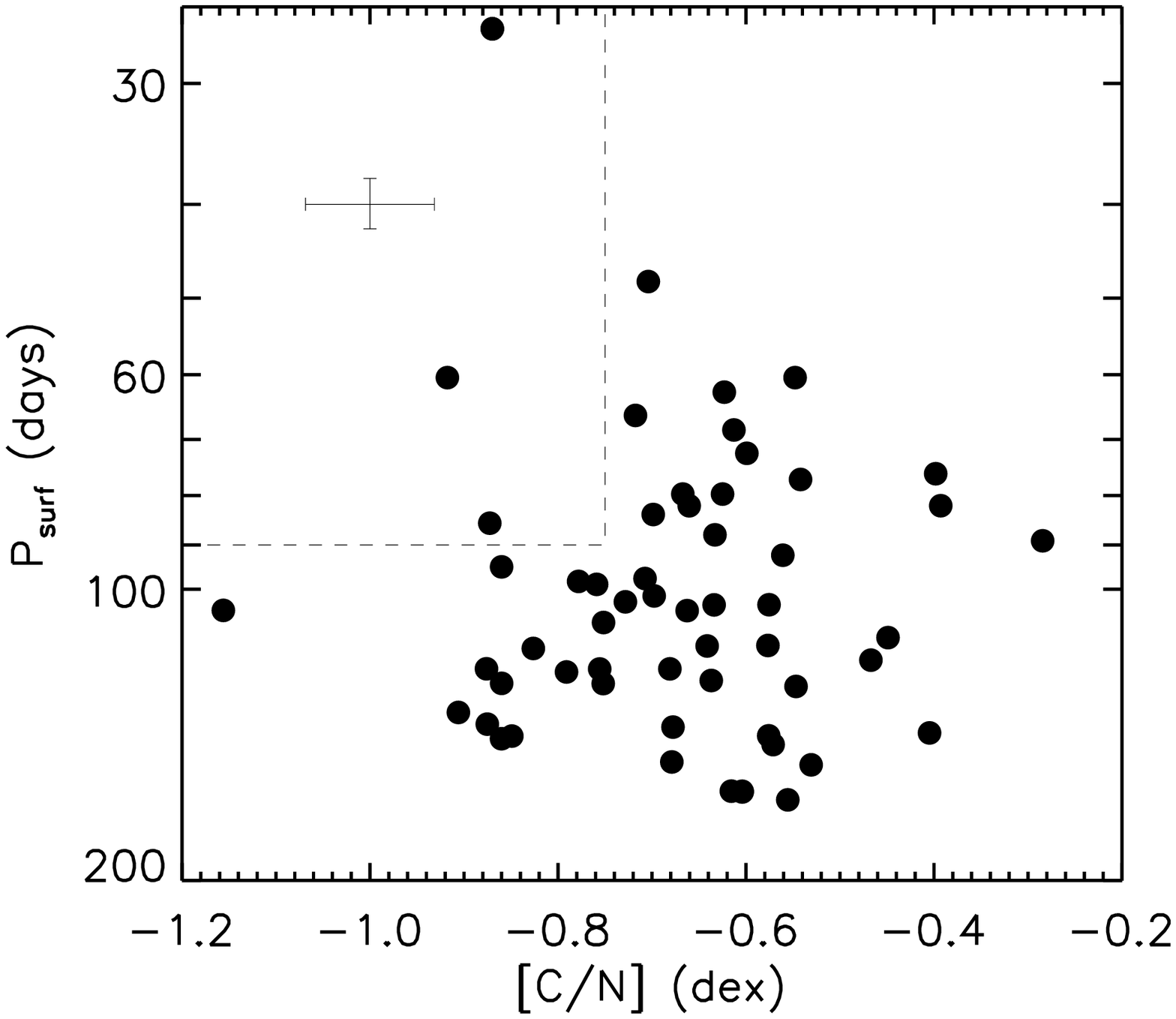}}
\subfigure{\includegraphics[width=8.cm, clip=true, trim=0.3in 0.2in 0.2in 0.5in]{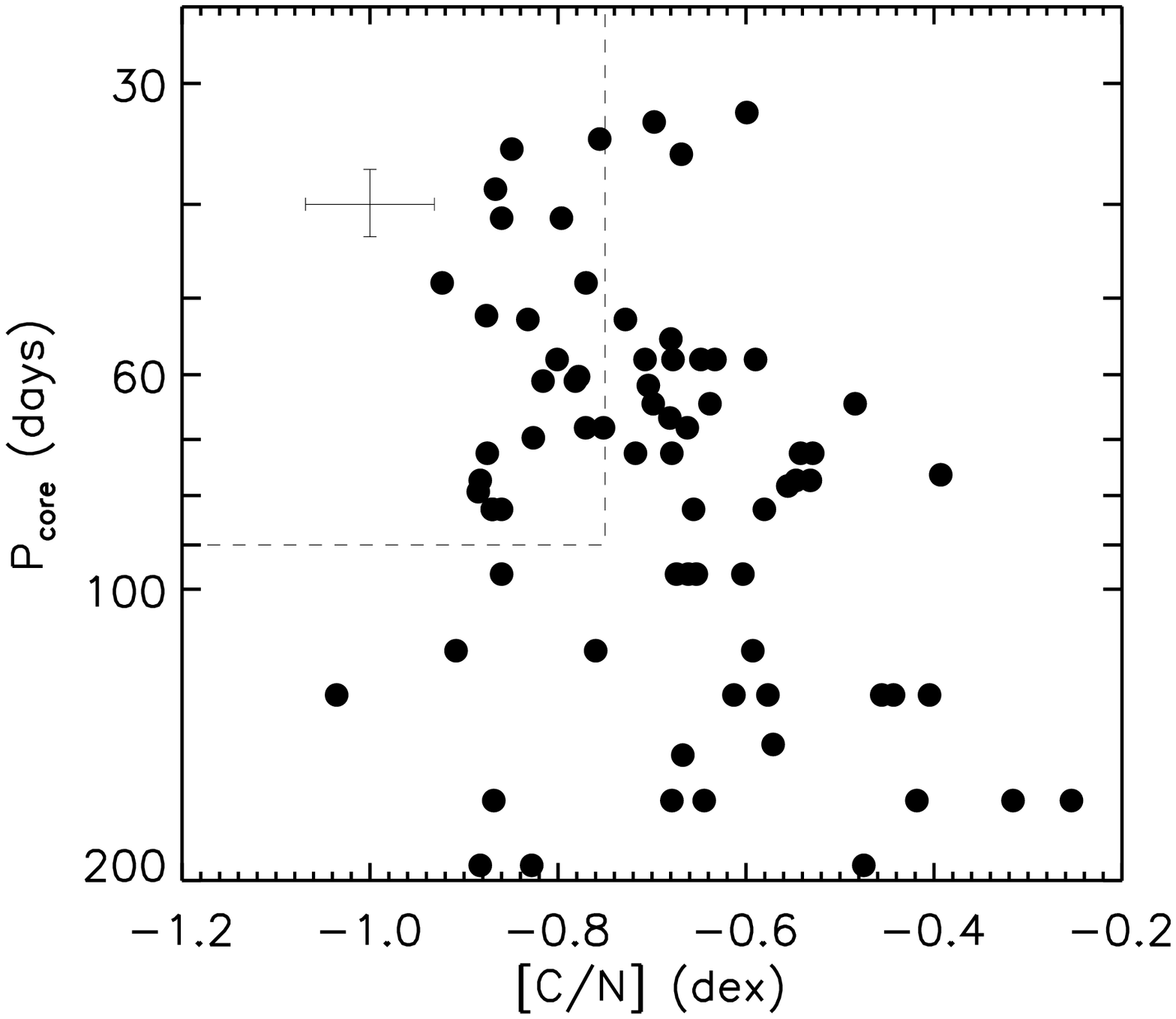}}
\caption{The locations (top) in mass-[C/N] space of the secondary clump stars with measured core rotations (black circles), and those where we are unable to measure core rotation rates (red xs).  We see no strong correlation between surface (middle) or core (bottom) rotation rate and [C/N], a mixing diagnostic, although we notice an interesting lack of fast rotating surfaces with low [C/N] values (dashed box), which could be driven by stellar interactions. }
\label{Fig:CNnon}
\end{center}
\end{figure}

The ratio of carbon to nitrogen, [C/N], at the surface of a red giant is considered a mixing diagnostic because it is significantly affected by how much nitrogen-rich material is dredged up from the stellar interior. \citet{Martig2016} and \citet{Ness2016} showed empirically that the [C/N] ratio is strongly correlated with stellar mass and therefore age. While this dependence of [C/N] on mass is significant for low-mass stars, we show in Figure \ref{Fig:CNnon} {(top)} that the trend flattens for stars above $\sim$2.0 \msun, such that all stars between 2.0 and 3.0 \msun\ have [C/N] of order $\sim-0.7$ dex, although there is still significant scatter. If this residual scatter is driven by rotational mixing, we might expect a correlation between [C/N] and either the surface (Figure \ref{Fig:CNnon}, {middle}) or core (Figure \ref{Fig:CNnon}, {bottom}) rotation rate. However, we see no such trends. We do however notice that the fraction of core rotation nondetections seems to be correlated with the [C/N] (Figure \ref{Fig:CNnon}, {top}), where stars with [C/N] higher than the average (less mixed) are more likely to not have measured core rotation rates,{ and a Kolmogorov\textendash Smirnov test indicates that this difference is significant at the {99} percent confidence level}. We suggest that this could indicate that there is some process that affects both the strength of the mixed modes and the surface composition. For example, the strong core magnetic fields suggested by \citet{Fuller2015} to suppress the mixed modes required to measure core rotation might also affect the mixing in these stars, or a history of stellar interactions could be affecting the surface chemistry and also the interior structure in a way that makes the core rotation hard to measure. We suggest that a closer look at these stars might help determine the cause of this correlation, and in turn reveal something about the mixing history of these stars.

We also notice a tentative lack of low [C/N] stars with fast rotating surfaces (P$_{surf}<90$ days). These periods are above the range predicted by our single star models for most of the secondary clump, and we suggest that some of the stars rotating this fast might have been spun up by a stellar interaction. If this is the case, then the lack of low [C/N] values in these stars could suggest (1) mass accretion, where these stars were less massive during the first dredge-up, and therefore did not reach such low [C/N] values, (2) the accretion of unmixed, high-carbon material, or (3) an inhibition of mixing during the merger, although we caution that this result is not strongly significant and requires a larger sample of stars where core rotation measurements were attempted. Additionally, the fact that this result only shows up in the surface rotation periods and not in the core rotation periods would suggest that stellar interactions do not or are slow to affect the core rotation rate, consistent with suggestions by \citet{Beck2018}.

\subsection{Core-Envelope Recoupling} \label{Sec:core-envelopecontrast}

There were {60} stars in our sample that had measured reliable surface rotation periods from \citet{Ceillier2017}. We particularly selected these stars because we wanted to use them to examine the core-surface contrast and how it evolves on the secondary clump. However, since only 2\% of red giants have detectable surface rotational modulation \citep{Ceillier2017}, and the amount of surface activity is often correlated with the stellar rotation rate \citep{Noyes1984}, we therefore need to test whether active stars were biased tracers of the core and surface rotation distributions.

In \citet{TayarPinsonneault2018} we demonstrated that modern angular momentum loss models can be used to successfully predict the surface rotation rates of stars in this evolutionary phase. For each of these cases on the following plots, we show the predicted rotation for a moderate rotator on the main sequence \citep[surface rotation rates of $\sim$150 \kms,][]{ZorecRoyer2012}, with ranges determined assuming main-sequence rotation velocities of 50 and 250 \kms, to bracket the observed range of main-sequence stars in the studied mass range. We note that, given the lack of information in literature on the range of rotation rates in intermediate-mass stars as a function of metallicity, we use the same initial rotation distributions for stars at all metallicities.

\begin{figure}[!htb]
\begin{center}
\includegraphics[width=9cm,clip=true, trim=0.5in 0in 0in 0in]{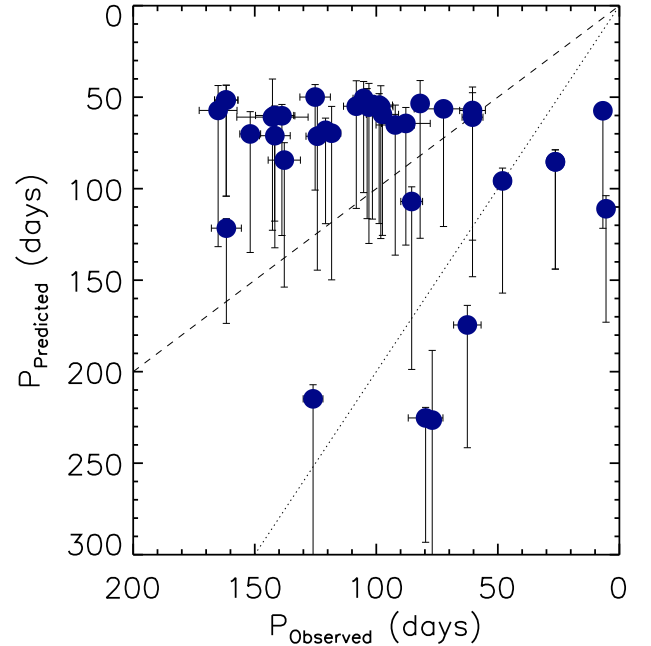}
\caption{Comparison between the surface rotation periods predicted from our theoretical models \citep{TayarPinsonneault2018} and those measured by \citet{Ceillier2017}. The dashed line represents perfect agreement, and the dotted line represents an alias of half the true period. While most of the periods are consistent within error bars, there are a number of stars with observed periods significantly shorter than the predicted period, something that could represent selection effects, binary interactions, aliasing in the period measurements, or errors in our theoretical models. }
\label{Fig:periodcompare}
\end{center}
\end{figure}

\begin{figure}[!htb]
\begin{center}
\includegraphics[width=9cm,clip=true, trim=0.5in 0in 0in 0in]{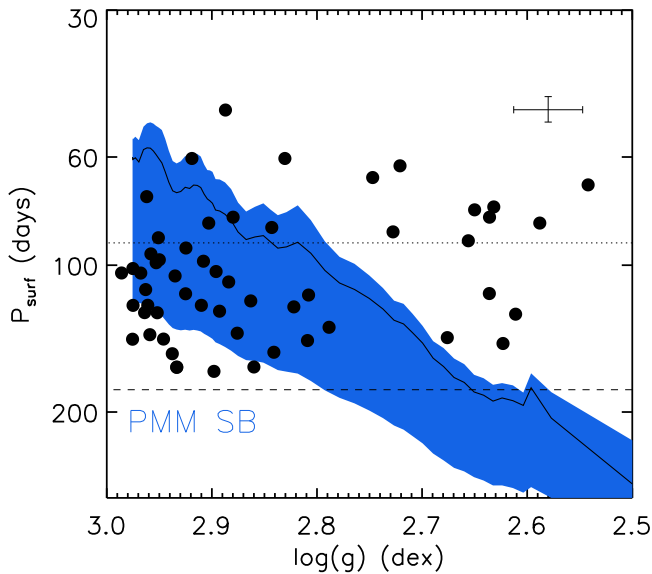}
\caption{Comparison between the measured surface rotation periods used in this work (black points) and the predicted evolution of surface rotation from \citet{TayarPinsonneault2018} (blue band). We have marked the periods corresponding to one (dotted line) and two (dashed line) \kepler\ quarters of data. The fact that the observed periods do not increase on average as the star expands suggests that our sample of measured surface rotation periods is not representative of the underlying population, likely due to the difficultly of measuring spot periods for slowly rotating, relatively inactive stars.}
\label{Fig:surftheory}
\end{center}
\end{figure}

\begin{figure}[!htb]
\begin{center}
\includegraphics[width=9cm,clip=true, trim=0.5in 0in 0in 0in]{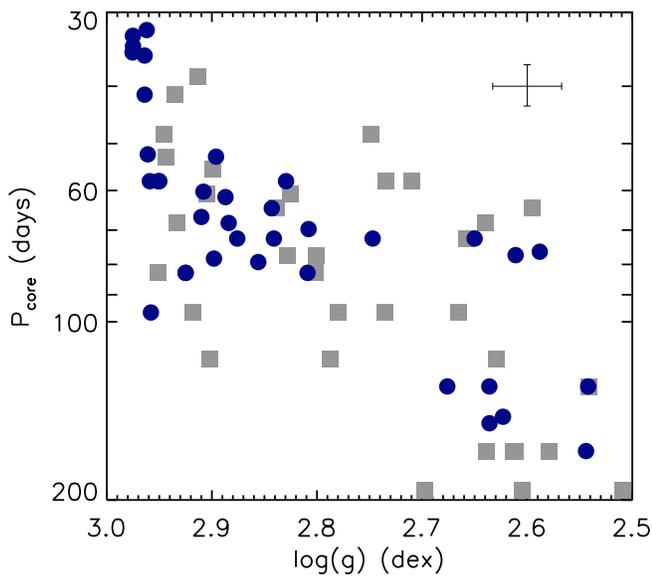}
\caption{Comparison between the core rotation rates for stars with measured surface rotation periods (blue circles) and those without (grey squares). Despite the biases in the measured surface rotation periods shown in previous plots, we see no indication of a difference in core rotation rates for active stars with measured surface rotation periods and similar inactive stars.}
\label{Fig:activity}
\end{center}
\end{figure}

We show in Figure \ref{Fig:periodcompare} that while the majority of observed periods are consistent within errors with our theoretical predictions, there is an interesting population of stars with observed periods substantially shorter than we would have predicted. These stars could be the descendants of interacting systems that gained angular momentum from a companion, but they could also be the result of measuring half the actual period, a common problem given the difficulty of measuring relatively small-amplitude, long-period signals through the systematic trends caused by the quarterly roll of the satellite. In addition, we show in Figure \ref{Fig:surftheory} that the lower envelope of the measured rotation rate distribution does not seem to evolve to longer period with larger radius, which  
contrasts with the predictions of angular momentum conservation. In addition, we note that the distribution in gravity of stars where surface rotation was measured is not consistent with our underlying sample. Together, these confirm that the active stars are indeed a biased tracer of the surface rotation distribution and should be used with caution. It also raises questions about whether their core rotation rates are also a biased tracer of the underlying distribution, and whether we should exclude them from the sample.
However, we show in Figure \ref{Fig:activity} that the active stars do not have significantly different core rotation rates than the inactive stars of similar gravity. We therefore tentatively conclude that while there might be a bias in the observed surface periods, there is not a strong correlation between the surface activity and core rotation in our sample.

\begin{figure}[!htb]
\begin{center}
\subfigure{\includegraphics[width=9cm,clip=true, trim=0.5in 0in 0in 0in]{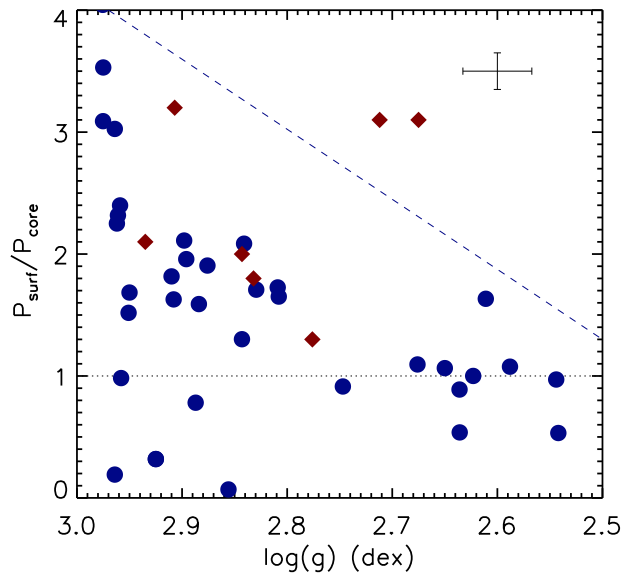}}
\subfigure{\includegraphics[width=9cm,clip=true, trim=0.5in 0in 0in 0in]{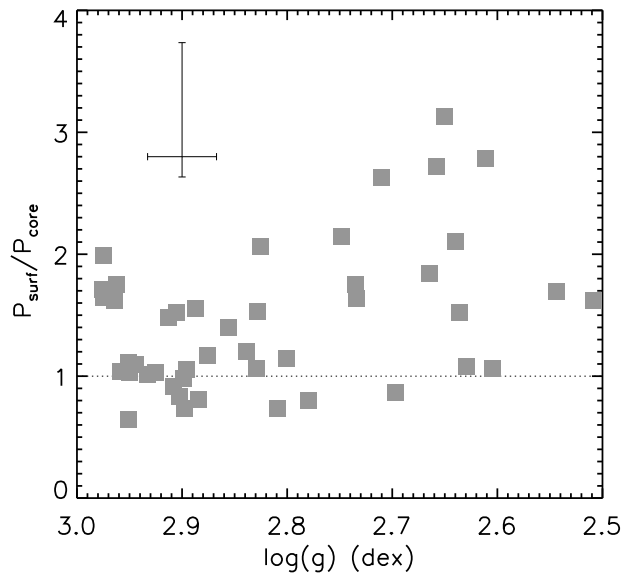}}
\caption{Comparison between the core and envelope rotation rates, with larger numbers representing faster cores and one representing a star rotating as a solid body. We show this ratio computed using measured surface rates rates in the top; those inferred from spot modulation are shown as blue circles \citep{Ceillier2017} and surface rates inferred from seismology are shown as red diamonds \citep{Deheuvels2015}. The difficulty of using spots to measure surface rotation periods longer than 200 days biases us against detecting ratios above the blue dashed line. In the bottom figure, we show core-envelope contrasts inferred using surface rotation rates inferred from stellar models are shown as grey squares \citep{TayarPinsonneault2018}, with error bars representing the range of intial rotation rates. While the data seems to indicate core-envelope recoupling, we must be careful of the selection effects in the stars where rotation can be measured, and it is interesting that the theoretical predictions seem to indicate approximately constant core-envelope contrast ratios or perhaps even core-envelope decoupling.}

\label{Fig:coreenv}
\end{center}
\end{figure}

In Figure \ref{Fig:coreenv} we show the ratio between the core and surface rotation rates for the 33 stars in our sample where both are measured (dark blue circles). We also show the ratio between the seismically measured core and envelope rotation rates from \citet{Deheuvels2015} (red diamonds). As previously discussed, however, these periods are likely to be biased tracers of the distribution. Therefore, 
in order to avoid bias from more evolved stars with longer surface rotation periods that were not detectable in \citet{Ceillier2017}, we predict a surface rotation rate and thus a core-envelope contrast for {the} stars in our sample (grey squares) by assuming these stars started out as moderate rotators on the main sequence. In addition to these observations,  solid-body rotation, a ratio of one, is marked as a dotted line. There seems to be some evidence in the data that the difference between the core and the envelope rotation is decreasing on the secondary clump as stars evolve to lower surface gravities, but this trend is not as significant when using theoretical priors on the surface rotation rates. In fact, the theoretical predictions show some preference for increasing core-envelope contrast as the star evolves to lower gravities. We therefore can not comment authoritatively on whether stars are decoupling or recoupling on the secondary clump and we encourage the collection of a larger unbiased sample of surface rotation rates for these stars, particularly the low-gravity stars.

\begin{figure*}[!htb]
\begin{minipage}{1.0\textwidth}
\begin{center}

\subfigure{\includegraphics[width=0.45\textwidth,clip=true, trim=0.5in 0in 0in 0in]{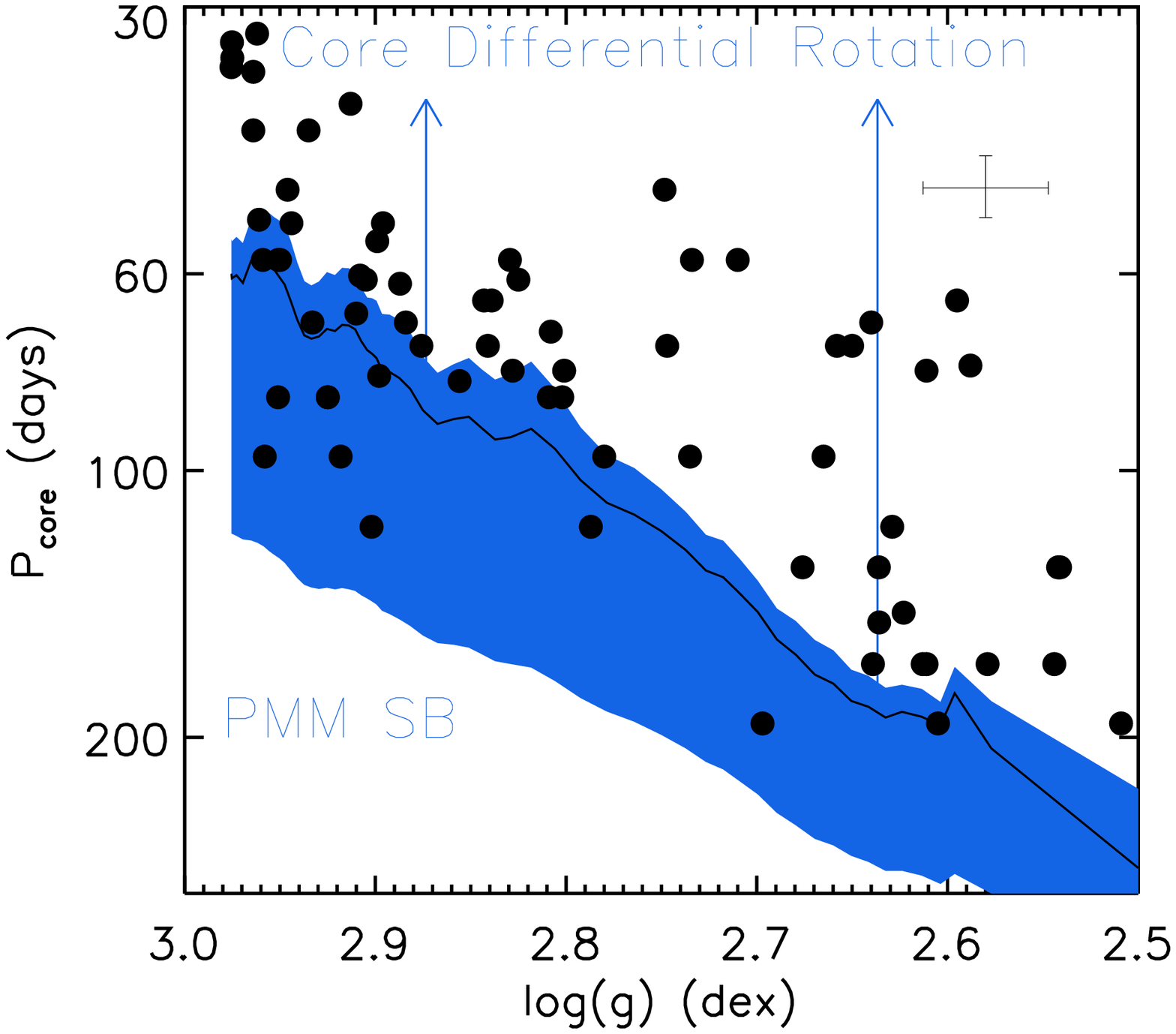}}
\subfigure{\includegraphics[width=0.45\textwidth,clip=true, trim=0.5in 0in 0in 0in]{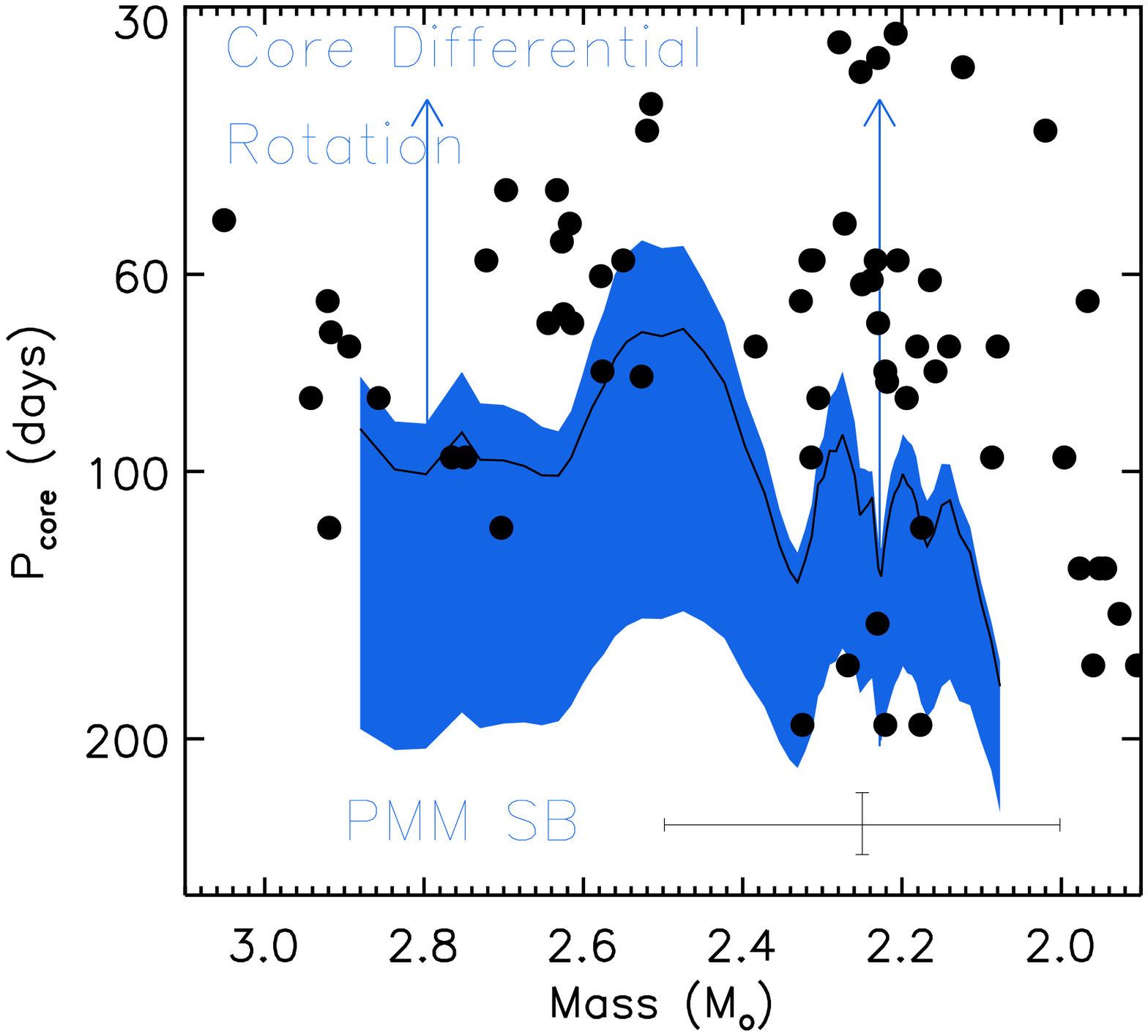}}
\subfigure{\includegraphics[width=0.45\textwidth,clip=true, trim=0.5in 0in 0in 0in]{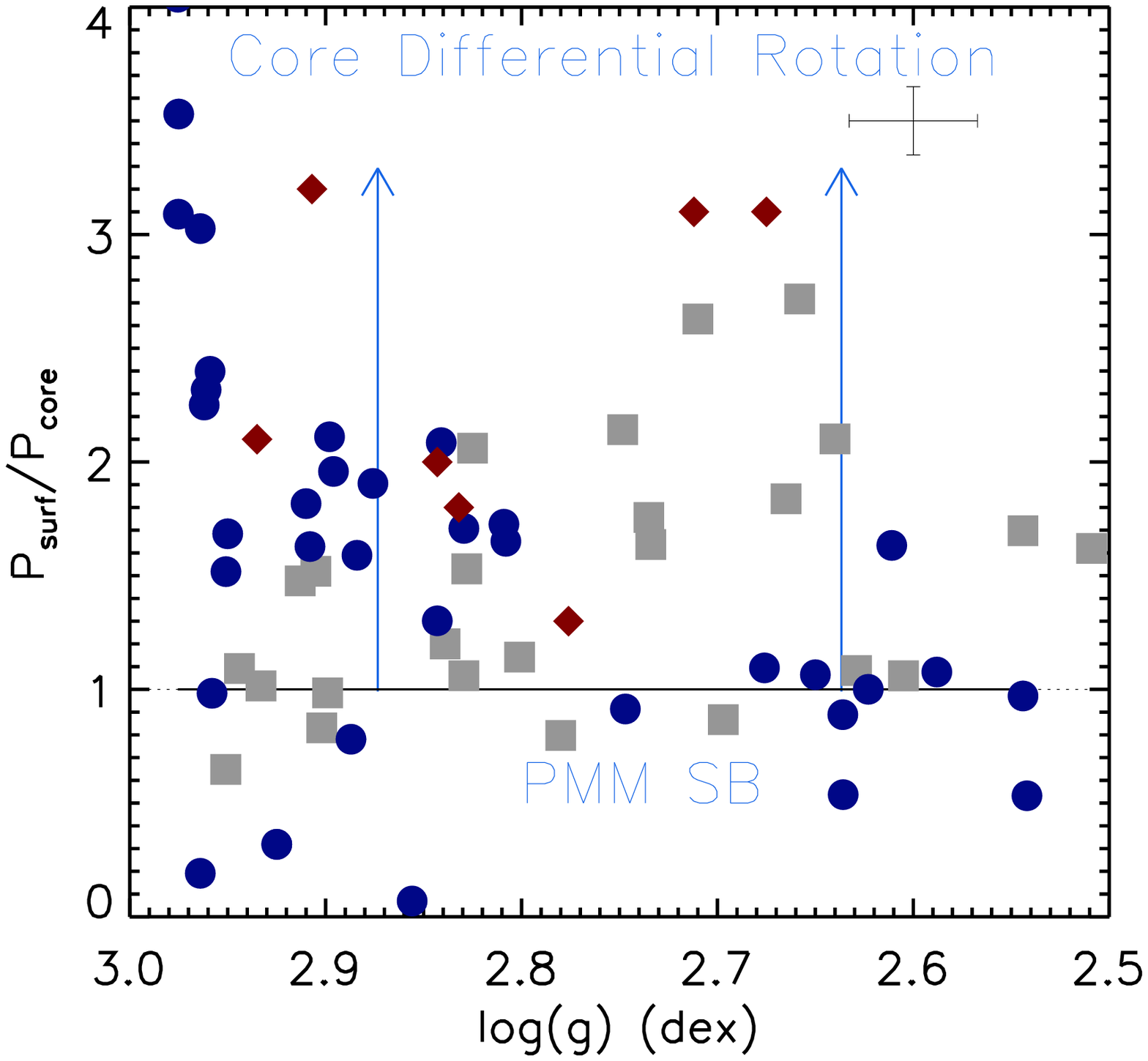}}
\subfigure{\includegraphics[width=0.45\textwidth,clip=true, trim=0.5in 0in 0in 0in]{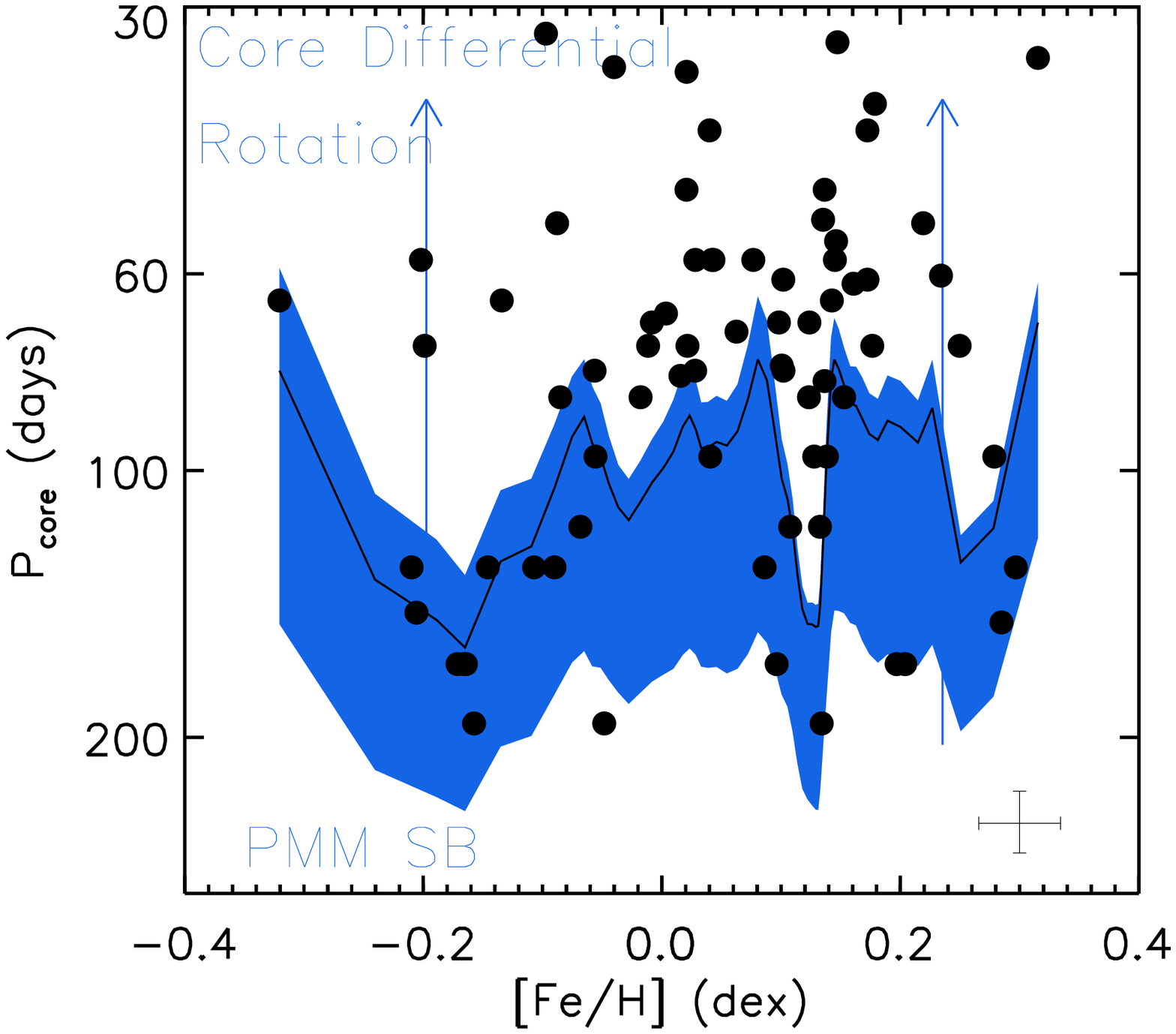}}

\caption{Comparison between our \textbf{solid body model} predictions and measured core rotation rates for the stars in our sample as a function of gravity (Upper Left), mass (Upper Right), metallicity (Lower Right). Lower Left: We show the predicted and observed ratio of surface to core rotation periods. In this case, a solid body model by definition always predicts a ratio of 1, and so that is represented by a solid line. In each panel, arrows indicate the direction differential rotation in the core would shift the predictions.  Characteristic error bars are shown in each panel.} 
\label{Fig:solidbody}
\end{center}
\end{minipage}
\end{figure*}

\begin{figure*}[!htb]
\begin{minipage}{1.0\textwidth}
\begin{center}

\subfigure{\includegraphics[width=0.45\textwidth,clip=true, trim=0.5in 0in 0in 0in]{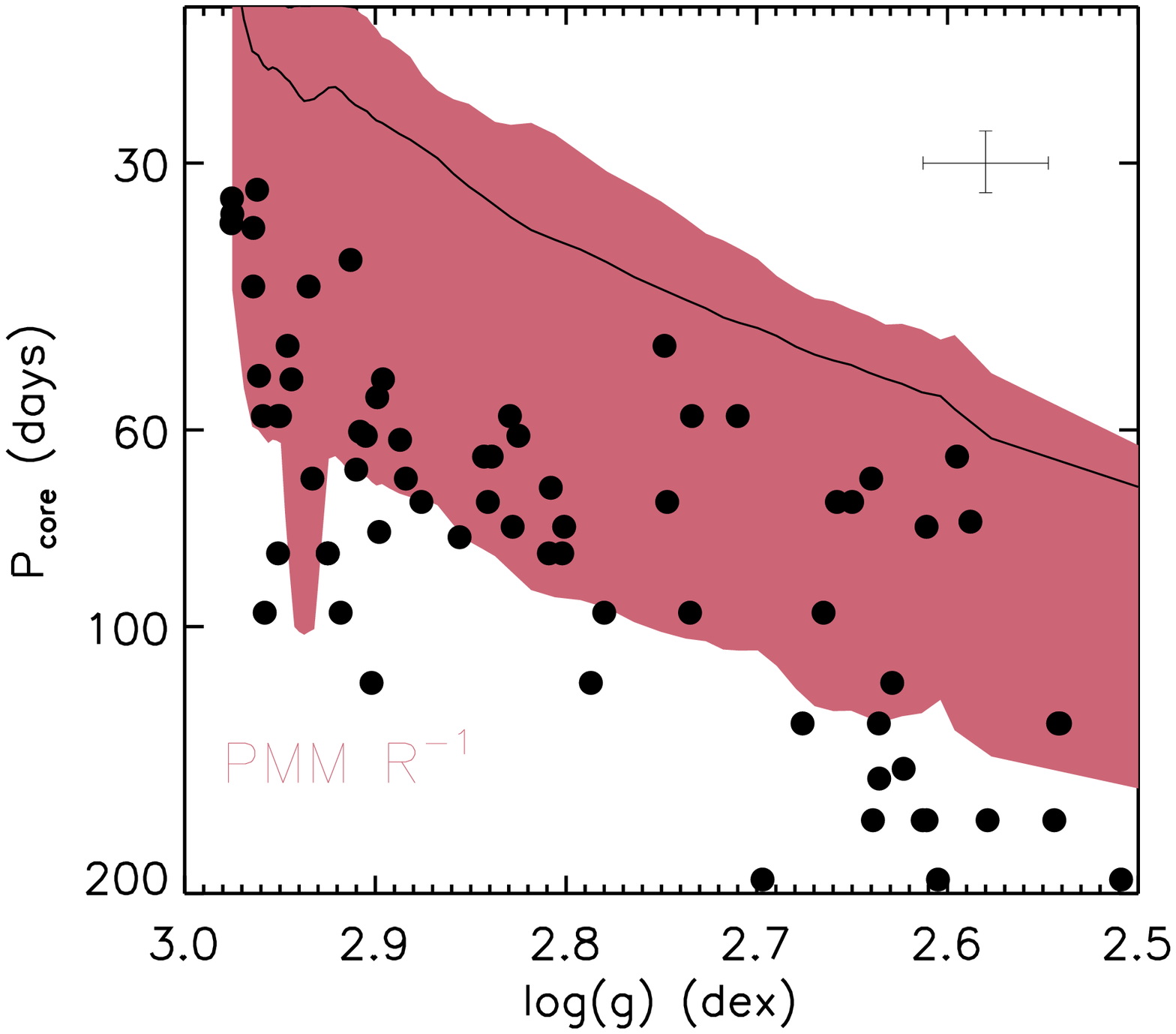}}
\subfigure{\includegraphics[width=0.45\textwidth,clip=true, trim=0.5in 0in 0in 0in]{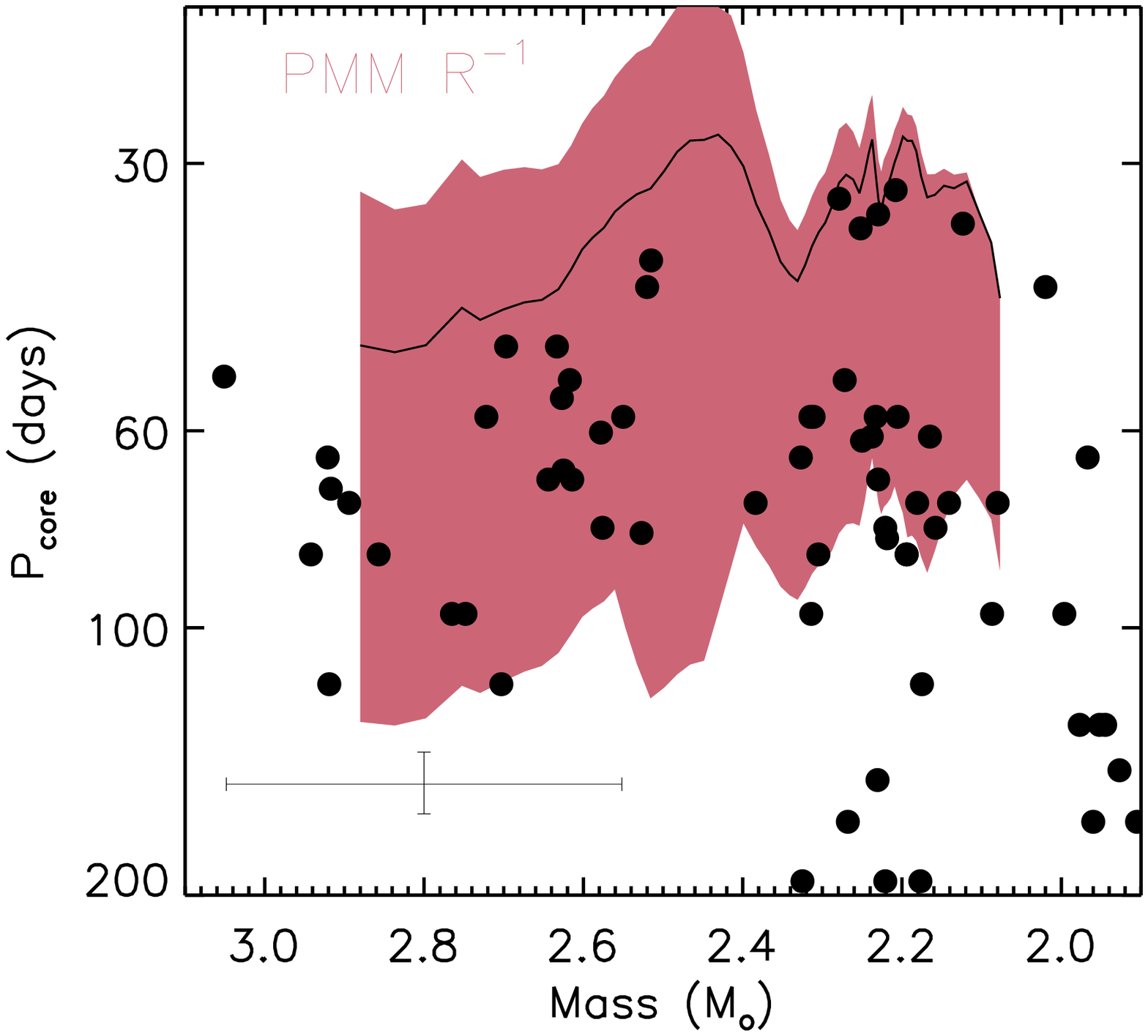}}
\subfigure{\includegraphics[width=0.45\textwidth,clip=true, trim=0.5in 0in 0in 0in]{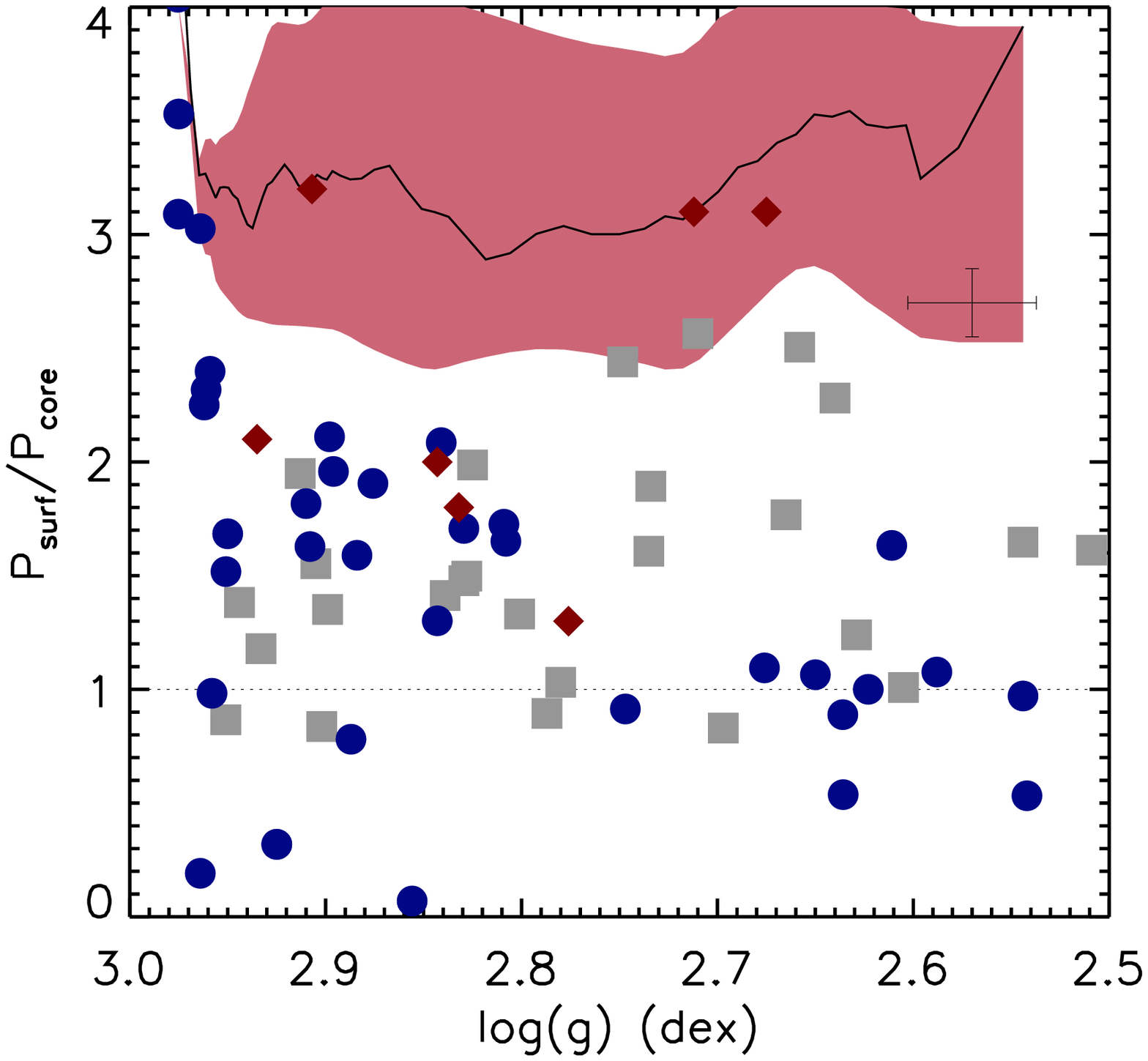}}
\subfigure{\includegraphics[width=0.45\textwidth,clip=true, trim=0.5in 0in 0in 0in]{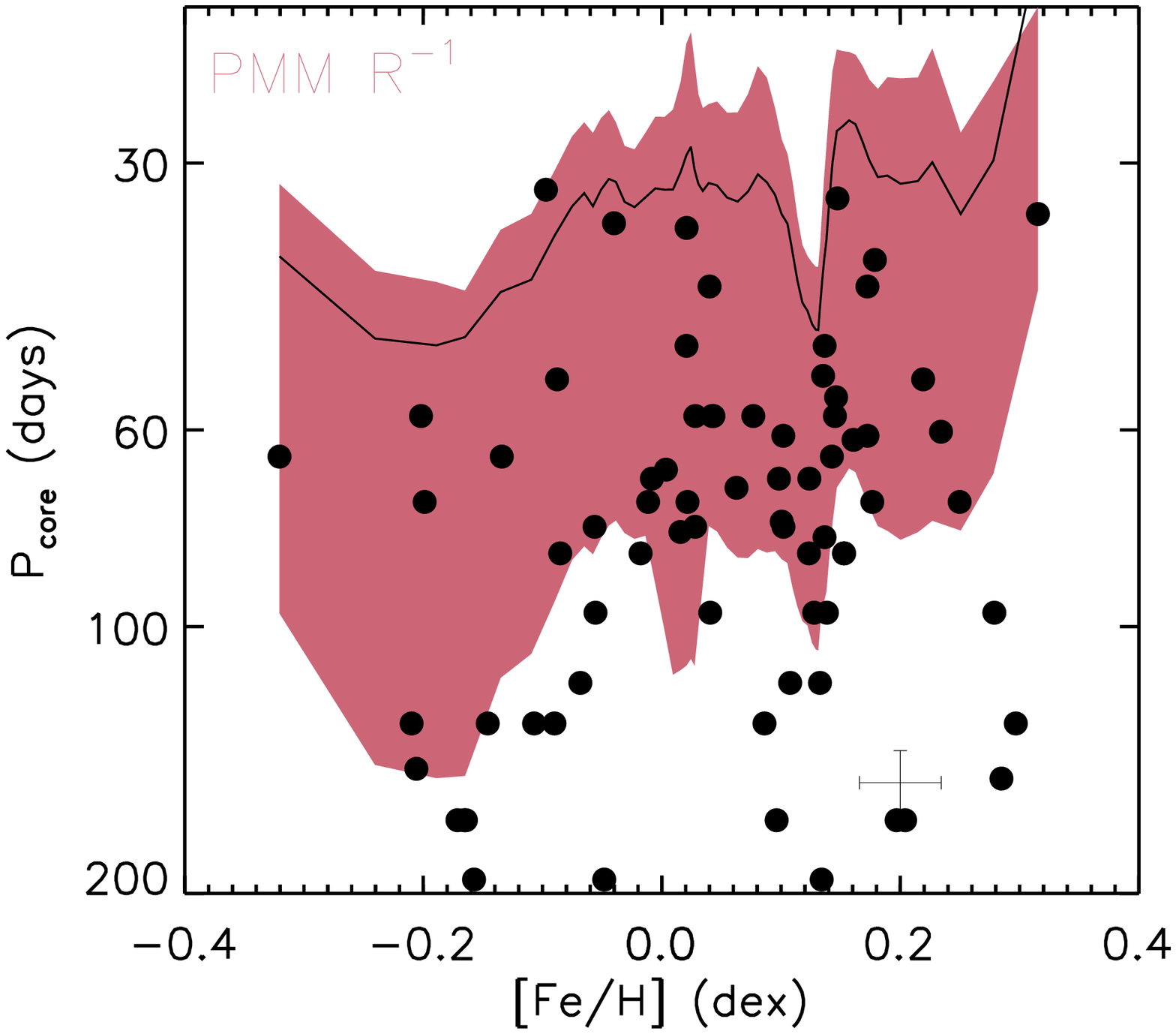}}
\caption{Comparison between our \textbf{differentially rotating surface convection zone model} and measured core rotation rates for the stars in our sample as a function of gravity (Upper Left), mass (Upper Right), metallicity (Lower Right). Lower Left: We show the predicted and observed ratio of surface to core rotation periods, with the range of values now representing the range of values expected for the ranges of masses and metallicities in our sample rather than the range of initial rotation periods. Characteristic error bars are shown in each panel.}
\label{Fig:modeling}
\end{center}
\end{minipage}
\end{figure*}

\section{Comparison With Models}
The empirical trends described above need to be placed in the context of stellar interiors models.  Secondary clump stars evolve significantly in surface gravity, in the sense that more evolved stars have larger radii and thus lower surface gravities. The moment of inertia of the star and envelope therefore increase significantly during evolution on the secondary clump. A significant reduction in envelope rotation is therefore required from angular momentum conservation, and could be enhanced from angular momentum loss. Because angular momentum loss is more efficient for rapid rotators, we expect to see a convergence in surface rotation rates for more evolved stars. For rigid internal rotation the core rotation trends would mirror the envelope ones. 

However, the moment of inertia of the core changes very little during the core He-burning phase, so if the core was uniformly rotating and detached from the envelope, we would expect little change in core rotation periods as a function of surface gravity.

Another natural scenario involves differential rotation in the radiative core, but not in the convective envelope.  In this case we would expect to see an intermediate pattern if there was effective internal angular momentum transport: a reduction in core rotation rates with decreased surface gravity. In the limit of full coupling, the core rates would converge to the surface ones.

If there was differential rotation in the convective envelope as well, there would be a similar overall pattern, with an important difference. If we make the reasonable assumption that the core rotation rate must be at least as high as the rate at the base of the surface convection zone, there should always be a minimum contrast between the core and surface rotation rates set by the differential rotation in the envelope itself.

We adopt the angular momentum loss models of \citet{TayarPinsonneault2018} for this section. The first model we explore is the solid body rotation model shown in Figure \ref{Fig:solidbody}. In this figure, we interpolate in our models to the properties of the observed stars in our sample in order to mitigate the effects of our heterogeneous sample selection. We then combine together and smooth the model predictions for the whole sample in order to predict an average core rotation rate as a function of surface gravity, mass, and metallicity, and we note that these predictions have structure because of the exact stars included in our sample. We also expect the range of surface rotation rates at the end of the main sequence for these stars to map to a range of predicted core rotation rates, since we assume that these stars rotate as solid bodies on the main sequence. We use the predictions for a slow rotating star ($\sim$50\kms) and a fast rotating star ($\sim$250\kms), representing the range observed by \citet{ZorecRoyer2012}, to create a colored band of possible rotation rates, since this uncertainty in initial rotation rates dominates over any uncertainties in the measured properties of the stars. {As indicated in Table \ref{Table:summary}, not all stars in the sample have predicted rotation rates. 
This was most common for stars whose final estimated mass was below 2~\msun, but there were also problems with stars at the highest surface gravities.  As shown in Figure \ref{Fig:SeisSel}, the maximum surface gravity of the clump is strongly mass dependent, and stars with observed surface rotation periods tended to cluster at higher gravities than the bulk of the sample. We are unsure if this clustering is fully explained by a selection effect, since periods are easiest to measure in the physically smallest stars, or if there is some other process such as mass transfer or observational errors causing this shift, but that exploration is outside the scope of the current work. However,
} in order to reduce the number of stars that fall outside of the range of our model grid, we have artificially shifted the surface gravities of the models by 0.05 dex. {This is roughly equivalent to changing the amount of convective overshooting in the models, and increases the estimated surface rotation periods by about 20 percent on average, but allows us to predict theoretical surface rotation periods for about 60 percent more stars.}  

In Figure \ref{Fig:solidbody}, we see that the model predictions tend to trace the structure of the observations but at rates slower than we observe. First, this strengthens our conclusion that the solid body models can be treated as a lower limit on the core rotation speed, and that in general the cores of these stars rotate faster than their envelopes. There are two important features derived from these models.  Although core rotation is reduced in an absolute sense, the rotation periods do not converge to a unique value, nor do they reach the surface rotation rates.  The coupling timescale is therefore longer than the core He burning lifetime, and stars retain a memory of their birth rotation rates.
As seen in Figure \ref{Fig:coreenv}, the direct core to surface ratios (top panel) show different trends than the ones predicted by evolutionary models (bottom panel of \ref{Fig:coreenv}, and Figure \ref{Fig:solidbody}).  We argue that this is likely to be a selection effect, because the long rotation periods expected from angular momentum conservation and loss in more evolved stars are difficult to measure. Direct studies of surface rotation in such stars, either with \vsini\ or asteroseismology, would be very helpful for clarifying this picture.

\begin{figure}[!htb]
\begin{center}
\includegraphics[width=9cm,clip=true, trim=0.5in 0in 0in 0in]{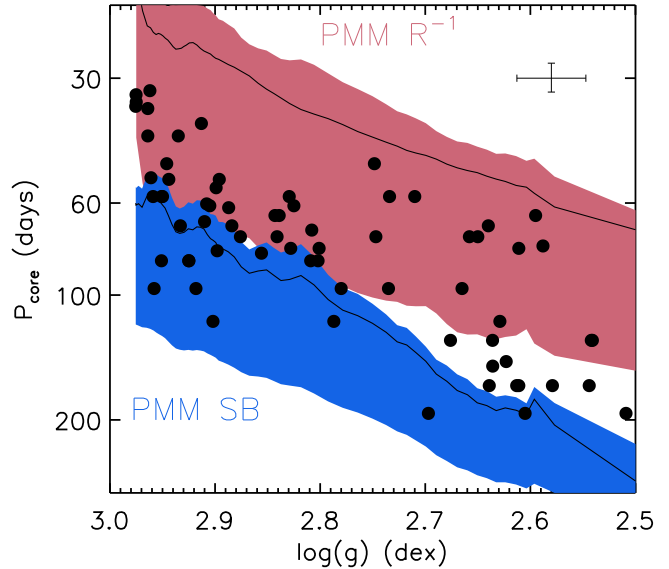}
\caption{The measured core rotation rates for the stars in our sample as a function of gravity compared to the predictions of our solid body model (blue) and our model with a moderately differentially convection zone (pink), showing that these models provide limits on the allowable amount of radial differential rotation in the surface convection zone.}

\label{Fig:bothmodels}
\end{center}
\end{figure}

In Figure \ref{Fig:modeling}, we show the consequences of assuming that the differential rotation is happening entirely in the surface convection zone at some moderate rate ($\Omega \propto R^{-1}$). In this case we again generally fit the trends of the observed rotation, but predict cores that rotate faster than the observations, suggesting that this case can be treated as an upper limit on the amount of radial differential rotation that can be permitted in the surface convection zone. In particular, the absence of differential rotation in some stars is very difficult to explain in a model where the interaction of rotation and convection always produces a base level of internal differential rotation. While we are able to place limits on the possible range of differential rotation in the convective envelope (see Figure \ref{Fig:bothmodels}), we choose not to fit a preferred value both because of the possibility of differential rotation in the radiative core and because the interplay between the angular momentum loss and the surface differential rotation makes the problem somewhat nonlinear, and thus difficult to constrain with the available grid of models.

Recent theoretical work using a diffusion coefficient approach \citep{denHartogh2019} reaches conclusions that are complementary to ours.  They found that reproducing the \citet{Deheuvels2015} trends required a relatively high diffusion coefficient during the secondary clump, but consistency with other data sets required much lower ones before and after.  This is consistent with the picture from our data, where stars arrive on the secondary clump with significant internal differential rotation (implying incomplete prior coupling), and experience some recoupling during the phase, and perhaps decouple somewhat near the end of the secondary clump. A second model by \citet{Fuller2019} predicts relative trends in core rotation rates with surface gravity that are perhaps slightly weaker than what we observe, although their lack of mass trends is consistent with our observations. We do however see a range of rotation rates at all surface gravities in the clump that at first seems inconsistent with their predictions, although a more detailed comparison of these theoretical models to our data would be an interesting exercise. {It has also been suggested that internal gravity waves generated by convection could transport the necessary amount of angular momentum \citep[e.g.][]{Aerts2019}, and the recent claimed detection of these waves in massive stars \citep[][but see also \citealt{Lecoanet2019}]{Bowman2019} is an encouraging step forward for being able to model exactly how much angular momentum they should transport.}

In summary, the observed picture illustrates that secondary clump stars are in a dynamic state.  They are not in equilibrium, either fully decoupled or fully coupled. If we adopt surface rotation periods from evolutionary models, there is both a clear need for coupling during the secondary clump phase and clear evidence that such coupling is incomplete.

The secondary clump is a powerful test of the theory of stellar structure, and a rare window on the internal rotation of massive stars.  Looking forward, the sample of core rotation measurements can be increased significantly once the full \textit{Kepler} sample is analyzed.  The measurement of surface rotation rates is challenging in these stars, but important, and our work indicates that significant progress on that front will be important. More precise mass data would also be helpful in looking for trends, as would be optical data for mixing diagnostics such as Li. Looking further ahead, stars in the TESS continuous viewing zone, and PLATO targets, can add new information as well to our understanding of angular momentum transport in stellar interiors.

\section{Summary}
\begin{itemize}[leftmargin=*]
    \item We measure core rotation periods for 72 intermediate-mass core-helium burning secondary clump stars.
    \item Core rotation periods decrease with surface gravity, model-independent evidence of significant angular momentum transport between the core and envelope.
    \item We see no trends in core rotation with mass or metallicity.
    \item A range of core rotation rates persist through the entirety of the secondary clump.
    \item Intermediate-mass stars with high [C/N] values, indicative of less mixing, are less likely to have measurable core rotation rates, suggesting a possible correlation between rotation and mixing histories.
    \item The distribution of measured surface rotation periods for these stars is likely not representative of the underlying distribution.
    \item We see tentative evidence for core-envelope recoupling at the beginning of the core-helium burning phase, as well as core-envelope decoupling in the later parts of this phase.
    \item A small amount of radial differential rotation is required in these stars.
    \item If that radial differential rotation is occuring in the surface convection zone, it must be weaker than $\Omega \sim R^{-1}$.
\end{itemize}

\begin{acknowledgements}
{We thank the referee for a helpful report.}
We thank Jennifer van Saders and Daniel Huber for helpful comments that improved the manuscript.
JT and MHP acknowledge support from NASA grant NNX15AF13G. PGB acknowledges the support of the Spanish Ministry of Economy and Competitiveness (MINECO) under the programme 'Juan de la Cierva' (IJCI-2015-26034). {The authors gratefully acknowledge support from NAWI Graz.} RAG acknowledges the support from the CNES. S.M. acknowledges support by the National Aeronautics and Space Administration under Grant NNX15AF13G, by the National Science Foundation grant AST-1411685 and the Ramon y Cajal fellowship number RYC-2015-17697. Support for this work was provided by NASA through the NASA Hubble Fellowship grant No.51424 awarded by the Space Telescope Science Institute, which is operated by the Association of Universities for Research in Astronomy, Inc., for NASA, under contract NAS5-26555.

Funding for the Sloan Digital Sky Survey IV has been provided by
the Alfred P. Sloan Foundation, the U.S. Department of Energy Office of
Science, and the Participating Institutions. SDSS-IV acknowledges
support and resources from the Center for High-Performance Computing at
the University of Utah. The SDSS web site is www.sdss.org.

SDSS-IV is managed by the Astrophysical Research Consortium for the 
Participating Institutions of the SDSS Collaboration including the 
Brazilian Participation Group, the Carnegie Institution for Science, 
Carnegie Mellon University, the Chilean Participation Group, the French Participation Group, Harvard-Smithsonian Center for Astrophysics, 
Instituto de Astrof\'isica de Canarias, The Johns Hopkins University, 
Kavli Institute for the Physics and Mathematics of the Universe (IPMU) / 
University of Tokyo, Lawrence Berkeley National Laboratory, 
Leibniz Institut f\"ur Astrophysik Potsdam (AIP),  
Max-Planck-Institut f\"ur Astronomie (MPIA Heidelberg), 
Max-Planck-Institut f\"ur Astrophysik (MPA Garching), 
Max-Planck-Institut f\"ur Extraterrestrische Physik (MPE), 
National Astronomical Observatory of China, New Mexico State University, 
New York University, University of Notre Dame, 
Observat\'ario Nacional / MCTI, The Ohio State University, 
Pennsylvania State University, Shanghai Astronomical Observatory, 
United Kingdom Participation Group,
Universidad Nacional Aut\'onoma de M\'exico, University of Arizona, 
University of Colorado Boulder, University of Oxford, University of Portsmouth, 
University of Utah, University of Virginia, University of Washington, University of Wisconsin, 
Vanderbilt University, and Yale University.

This paper includes data collected by the Kepler mission and obtained from the MAST data archive at the Space Telescope Science Institute (STScI). Funding for the Kepler mission is provided by the NASA Science Mission Directorate. STScI is operated by the Association of Universities for Research in Astronomy, Inc., under NASA contract NAS 5–26555.

\end{acknowledgements}

\bibliographystyle{apj} 
\bibliography{ms}

\end{document}